# Asymptotic radiative transfer

## A C Selden*

ABSTRACT

Asymptotic radiative transfer (ART), like diffusion theory, assumes the angular intensity distribution incident at a boundary is identical with that in the depth of the scattering medium. However, the asymptotic intensity profile and accurate attenuation coefficient corresponding to the scattering phase function are used in preference to the P1 approximations of diffusion theory, thereby extending the range of asymptotic radiative transfer to the lowest particle scattering albedoes. Asymptotic calculations for scattering media with refracting boundaries are compared with diffusion theory and accurate radiative transfer data to illustrate this. The error involved can be further reduced by the addition of a suitable boundary transient.

*adrian_selden@yahoo.com

# Asymptotic theory

## Introduction

The asymptotic approach employs the partial fluxes $J_+$, $J_-$ obtained by integrating the angular intensity $I(\mu)$ over the forward and backward hemispheres [1, 2], where $\mu$ is the direction cosine with respect to the symmetry axis. In contrast to diffusion analysis, which assumes a mild cosine intensity variation and approximate diffuse attenuation coefficient [3], the asymptotic intensity distribution $I_{as}(\mu)$ valid in the depth of the scattering medium and exact attenuation coefficient (least eigenvalue $\lambda_0$) corresponding to the scattering phase function $p(\mu)$ are used here [4]. The asymptotic calculations yield more accurate results than diffusion theory, they minimize the number of boundary equations compared with other methods, thereby reducing the numerical effort in multilayer calculations, and are applicable over a broad range of scattering albedo $\varpi \in [0, 1]$.

## Partial fluxes

The partial fluxes $J_+$, $J_-$ are defined as [1, 2]

$$J_+ = \tfrac{1}{2} \int_0^1 I(\mu)\mu \, d\mu \qquad (1a)$$

$$J_- = \tfrac{1}{2} \int_0^1 I(-\mu)\mu \, d\mu \qquad (1b)$$

Here $I(\mu)$ corresponds to the asymptotic angular intensity distribution $I_{as}(\mu) \Rightarrow g(\mu, \lambda)$, the solution of the characteristic equation of radiative transfer valid in the depth of the scattering medium [Appendix A]. The asymptotic method proceeds by formulating the boundary equations in terms of the partial fluxes, yielding integrated boundary equations (IBEs) for the problem posed.

## Integrated boundary equations

At a specularly reflecting boundary, the forward intensity $I(\mu>0)$ equates to the reflected (backward) intensity $I(\mu<0)$ viz.

$$I(\mu) = R(\mu) I(-\mu) \qquad (2a)$$

and for a diffusely reflecting boundary

$$I(\mu) = R(\mu) J_- \qquad (2b)$$

where $R(\mu)$ is the boundary reflectance, which in general is a function of $\mu$. At a refracting boundary, $R(\mu)$ is given by the Fresnel reflectance function for unpolarised light [1]

$$R(\mu) = \frac{1}{2}\left[\left(\frac{\mu - n\mu_0}{\mu + n\mu_0}\right)^2 + \left(\frac{\mu_0 - n\mu}{\mu_0 + n\mu}\right)^2\right] \qquad \mu \geq \mu_c \qquad (3)$$

and $R(\mu) \equiv 1$ for $\mu \leq \mu_c$, the cosine of the critical angle: $\mu_c^2 = 1-1/n^2$ for refractive index n. Clearly, the reflected intensity profile $I(\mu>0)$ has a 'hole' for $\mu \in |\mu_c,1|$, corresponding to low reflectance within the cone defined by the critical angle (Fig 1). Multiplying eqns (2a, 2b) by $\mu d\mu$ and integrating, we obtain the integrated boundary equations (IBEs) for specular and diffuse reflection respectively

$$J_+ = \tfrac{1}{2} \int_0^1 I(\mu)\mu d\mu = \tfrac{1}{2} \int_0^1 R(\mu)I(-\mu)\mu d\mu$$
$$J_+ = \tfrac{1}{2} \int_0^1 I(\mu)\mu d\mu = J_- \int_0^1 R(\mu)\mu d\mu$$
(4a,b)

**Illuminated half space**

At the boundary of a refracting medium illuminated by a surface source $S(\mu_0)$, the internal forward intensity $I(\mu)$ equates to the specularly reflected backward intensity $R(\mu)I(-\mu)$ plus the transmitted source intensity concentrated by a factor $n^2$

$$I(\mu) = R(\mu)I(-\mu) + n^2[1-R(\mu_0)]S(\mu_0) \qquad (5)$$

where the interior and exterior cosines $\mu$, $\mu_0$ are related via Snell's law viz.[1]

$$1-\mu_0^2 = n^2(1-\mu^2)$$

whence $\qquad \mu_0 d\mu_0 = n^2 \mu d\mu \qquad$ (6 a,b)

Thus the integrated boundary equaton (IBE) for the asymptotic intensity $I_{as}(\mu) \Rightarrow A_0 g(\mu, \lambda)$ is

$$A_0 \int_0^1 [g(\mu,\lambda) - R(\mu)g(-\mu,\lambda)]\mu d\mu = \int_0^1 [1-R(\mu_0)]S(\mu_0)\mu_0 d\mu_0 \qquad (7)$$

enabling the coefficient $A_0$ and half-space albedo A* to be determined viz.

$$A^* = A_0 \int_{\mu_c}^1 [1-R(\mu)]g(-\mu,\lambda)\mu d\mu \qquad (8)$$

which can be written symbolically as

$$A^* = \frac{S(n)F(n,\lambda)}{C(n,\lambda)} \qquad (9)$$

with $S(n)$, $F(n,\lambda)$, $C(n,\lambda)$ defined in eqns (12), (14) below

**Dielectric slab**

For a dielectric slab of width d = 2a, the asymptotic intensity may be written as

$$I(x,\mu) = A e^{-\lambda x} g(\mu,\lambda) + B e^{\lambda x} g(-\mu,\lambda) \qquad x \in (-a,a) \qquad (10)$$

representing the intensity as the sum of the forward ($\mu>0$) and backward ($\mu<0$) asymptotic angular intensities $g(\mu, \lambda)$, $g(-\mu, \lambda)$ with coefficients A and B, where $\lambda$ is the attenuation coefficient. Applying boundary equations (2) and (5), multiplying by $\mu d\mu$ and integrating

over $\mu \in |0,1|$ we obtain two IBEs (for x = a, −a)

$$Ae^{\lambda a}C(n,\lambda) + Be^{-\lambda a}D(n,\lambda) = S(n)$$
$$Ae^{-\lambda a}D(n,\lambda) + Be^{\lambda a}C(n,\lambda) = 0 \quad (11a,b)$$

which determine the unknown coefficients A, B in terms of source strength $S_0$ and attenuation parameter $\lambda a$, where

$$C(n,\lambda) = \int_0^1 [g(\mu,\lambda) - R(\mu)g(-\mu,\lambda)]\mu d\mu$$
$$D(n,\lambda) = \int_0^1 [g(-\mu,\lambda) - R(\mu)g(\mu,\lambda)]\mu d\mu \quad (12a,b,c)$$
$$S(n) = \int_0^1 [1 - R(\mu_0)]S(\mu_0)\mu_0 d\mu_0$$

whence the slab albedo A* and diffuse transmittance T* are given by

$$A^* = \frac{S(n)[C(n,\lambda)F(n,\lambda) - D(n,\lambda)E(n,\lambda)\exp(-4\lambda a)]}{C^2(n,\lambda) - D^2(n,\lambda)\exp(-4\lambda a)}$$
$$T^* = \frac{S(n)\exp(-2\lambda a)[C(n,\lambda)E(n,\lambda) - D(n,\lambda)F(n,\lambda)]}{C^2(n,\lambda) - D^2(n,\lambda)\exp(-4\lambda a)} \quad (13a,b)$$

where
$$E(n,\lambda) = \int_{\mu_c}^1 [1 - R(\mu)]g(\mu,\lambda)\mu d\mu$$
$$F(n,\lambda) = \int_{\mu_c}^1 [1 - R(\mu)]g(-\mu,\lambda)\mu d\mu \quad (14a,b)$$

**Interface**

At an interface between two dielectric scattering media with refractive index ratio $n = n_1/n_2$, the intensity leaving the interface equates to the reflected incident intensity plus the intensity transmitted (and refracted) by the adjoining medium viz.

$$I(\mu) = R(\mu)I(-\mu) + n^2[1 - R(\mu')]I(\mu')$$
$$I(-\mu') = R(\mu')I(\mu') + [1 - R(\mu)]I(-\mu)/n^2 \quad (15a,b)$$

where $\mu$, $\mu'$ are the cosines in the adjoining media, related as in eq (6a).

The IBEs are formed as above, with I(x, $\mu$), I(x′, $\mu'$) expressed as in eqn (10), with coefficients $A_1$, $B_1$, $A_2$, $B_2$. The IBEs are expressed in matrix form as

$$\begin{bmatrix} D(n_1,\lambda_1) & C(n_1,\lambda_1) \\ E(n_1,\lambda_1) & F(n_1,\lambda_1) \end{bmatrix} \begin{bmatrix} A_1 \\ B_1 \end{bmatrix} = \begin{bmatrix} F(n_2,\lambda_2) & E(n_2,\lambda_2) \\ C(n_2,\lambda_2) & D(n_2,\lambda_2) \end{bmatrix} \begin{bmatrix} A_2 \\ B_2 \end{bmatrix} \quad (16)$$

with C(n,$\lambda$), D(n,$\lambda$), E(n,$\lambda$), F(n,$\lambda$) defined as above (eqns (12a,b,c), (14a,b)).

A sample calculation of the flux discontinuity at an interface between two half-spaces with uniform sources is given in Appendix B. Asymptotic and diffusion results are compared with radiative transfer (RT) in Fig 2. Tables of C(n,$\lambda$), D(n,$\lambda$), E(n,$\lambda$), F(n,$\lambda$) vs $\varpi$, with analytic solutions of eqns (15a,b) and their diffusion counterparts, are presented in Appendix C.

**Double layer**

The IBEs for a double layer, consisting of two free boundaries plus an interface, are formulated by combining boundary eqns (11a,b) and (15a,b), yielding four IBEs for the four intensity coefficients $A_1$, $B_1$, $A_2$, $B_2$, written in matrix form as **M=AS**, where

$$M = \begin{bmatrix} C(n_1,\lambda_1)e^{\lambda_1 a_1} & D(n_1,\lambda_1)e^{-\lambda_1 a_1} & 0 & 0 \\ 0 & 0 & D(n_2,\lambda_2)e^{\lambda_2 a_2} & C(n_2,\lambda_2)e^{-\lambda_2 a_2} \\ D(n_{12},\lambda_1)e^{\lambda_1 a_1} & C(n_{12},\lambda_1)e^{-\lambda_1 a_1} & -F(n_{21},\lambda_2)e^{\lambda_2 a_2} & -E(n_{21},\lambda_2)e^{-\lambda_2 a_2} \\ E(n_{12},\lambda_1)e^{\lambda_1 a_1} & F(n_{12},\lambda_1)e^{-\lambda_1 a_1} & -C(n_{21},\lambda_2)e^{\lambda_2 a_2} & -D(n_{21},\lambda_2)e^{-\lambda_2 a_2} \end{bmatrix} \quad A = \begin{bmatrix} A_1 \\ B_1 \\ A_2 \\ B_2 \end{bmatrix} \quad S = \begin{bmatrix} S_0 \\ 0 \\ 0 \\ 0 \end{bmatrix} \quad (17)$$

**Multilayers**

The interface eqns (15a,b) can be extended to multilayer media, the IBEs at each interface relating the intensity coefficients $A_k$, $B_k$ for each layer to those of the adjacent layer. These are solved sequentially via matrix inversion (Appendix C) to fit external boundary conditions e.g. surface sources or free boundaries. Thus for each pair of adjacent layers (k, k+1)

$$\mathbf{M_k A'_k} = \mathbf{M_{k+1} A''_{k+1}} \qquad (18a)$$

where

$$\mathbf{A'_k} = \begin{bmatrix} A_k\, e^{-\lambda_k a_k} \\ B_k\, e^{\lambda_k a_k} \end{bmatrix}$$

$$\mathbf{A''_{k+1}} = \begin{bmatrix} A_{k+1}\, e^{\lambda_{k+1} a_{k+1}} \\ B_{k+1}\, e^{-\lambda_{k+1} a_{k+1}} \end{bmatrix} \qquad (18b)$$

and

$$\mathbf{M_k} = \begin{bmatrix} D(n_{k,k+1},\lambda_k) & C(n_{k,k+1},\lambda_k) \\ E(n_{k,k+1},\lambda_k) & F(n_{k,k+1},\lambda_k) \end{bmatrix}$$

$$\mathbf{M_{k+1}} = \begin{bmatrix} F(n_{k+1,k},\lambda_{k+1}) & E(n_{k+1,k},\lambda_{k+1}) \\ C(n_{k+1,k},\lambda_{k+1}) & D(n_{k+1,k},\lambda_{k+1}) \end{bmatrix} \qquad (18c)$$

**Boundary transient**

The accuracy of asymptotic theory can be enhanced by the addition of a boundary transient, as in the K-integral method [5]. Thus, expressing the boundary intensity $I(0, \mu)$ as the sum of the asymptotic intensity $I_{as}(0, \mu) \Rightarrow g(\mu, \lambda)$ and a boundary transient $\psi(\mu)$

$$I(0, \mu) = g(\mu, \lambda) + \psi(\mu) \qquad (19)$$

the relative magnitudes of $\psi(\mu)$ and $g(\mu,\lambda)$ being determined by taking two moments of the boundary equations with weight functions $g_i(\mu)$ (i=1,2) [5]. Further analysis and results for a half-space are presented in the Supplement.

**Diffusion theory**

Diffusion theory employs the scalar flux (flux density) φ and vector flux J = –D∇φ (where D is the diffusion coefficient), defined as follows [6]

$$\varphi = \tfrac{1}{2}\int_{-1}^{1} I(\mu) d\mu$$
$$J = \tfrac{1}{2}\int_{-1}^{1} I(\mu) \mu d\mu \qquad (20)$$

Expanding the angular dependence of φ at the boundary to first order in μ

$$\varphi(0,\mu) = \tfrac{1}{2}\varphi_0(0) + \tfrac{3}{2}\mu\varphi_1(0)$$
$$\varphi_1(0) = -D\varphi_0'(0) \qquad (21)$$

applying boundary eq (2a), multiplying by μ and integrating, we find [7]

$$\varphi_0(0) = \lambda_s \varphi'_0(0) \qquad (22)$$

$$\lambda_s = \frac{2}{3}\frac{[1+3R_2]}{[1-2R_1]} \qquad (23)$$

Here $\lambda_s$ is the linear extrapolation distance beyond the boundary, where $\varphi_0(x) \Rightarrow 0$, and $R_1$, $R_2$ are defined by

$$R_n = \int_0^1 R(\mu)\mu^n d\mu \qquad (24)$$

Similarly, RT analysis yields a linear extrapolation distance d = d(n,κ), dependent on n and extinction coefficient κ [1, 7]. Inserting eq (21) in boundary eq (5) and integrating, we obtain an analytic expression for the albedo of a half-space with refracting boundary [8]

$$A^* = \bar{q}_s \frac{1 - 2R_1 - 2D\kappa(1-3R_2)}{1 - 2R_1 + 2D\kappa(1+3R_2)} \qquad (25)$$

with source term

$$\bar{q}_s = \int_0^1 q(\mu)[1-R(\mu)]\mu d\mu \qquad (26)$$

and [9, 10]

$$D\kappa = (1-\varpi)/\kappa \qquad (27)$$

where κ is the extinction coefficient and D the diffusion coefficient. For a non-refracting boundary (n=1) $R_1$, $R_2 \equiv 0$ and eqn (25) reduces to

$$A^* = \bar{q}_s \frac{1-2D\kappa}{1+2D\kappa} \qquad (28)$$

The diffusion analysis is readily extended to a finite layer i.e. slab geometry viz.

$$\varphi(x) = Ae^{-\kappa x} + Be^{\kappa x} \qquad x \in |-a, a| \qquad (29)$$

where 2a is the slab width. Applying boundary equations (2), (5) we find

$$(1-2R_1)(Ae^{\kappa a}+Be^{-\kappa a})+2D\kappa(1+3R_2)(Ae^{\kappa a}-Be^{-\kappa a})=4\overline{q_s}$$
$$(1-2R_1)(Ae^{-\kappa a}+Be^{\kappa a})+2D\kappa(1+3R_2)(Ae^{-\kappa a}-Be^{\kappa a})=0 \qquad (30a,b)$$

Solving for A, B yields the boundary fluxes φ(a), φ(−a) for determining the slab albedo A* and diffuse transmittance T* viz.

$$A^* = \overline{q_s}\frac{-[\beta_1-(\alpha_2/\alpha_1)\beta_2\exp(-4\kappa a)]}{\alpha_1[1-(\alpha_2/\alpha_1)^2\exp(-4\kappa a)]}$$

$$T^* = \overline{q_s}\frac{-[\beta_2-(\alpha_2/\alpha_1)\beta_1]\exp(-2\kappa a)}{\alpha_1[1-(\alpha_2/\alpha_1)^2\exp(-4\kappa a)]} \qquad (31a,b)$$

where

$$\alpha_1 = 1-2R_1+2D\kappa[1+3R_2]$$
$$\alpha_2 = 1-2R_1-2D\kappa[1+3R_2]$$
$$\beta_1 = 1-2R_1-2D\kappa[1-3R_2]$$
$$\beta_2 = 1-2R_1+2D\kappa[1-3R_2] \qquad (32a,b,c,d)$$

The diffusion boundary equations at an interface between two refractive scattering media are

$$\varphi_2 = n^2\varphi_1+C_0(n_{21})J_{1,2}$$
$$J_{1,2} = J_1 = J_2 \qquad (33a,b)$$

where $C_0(n_{21})$ is a monotonic function of refractive index ratio $n_{21} = n_2/n_1$ [1, 5]. Dividing by $\varphi_1$ we find

$$\varphi_2/\varphi_1 = n^2+C_0(n_{21})<\mu_1> \qquad (34)$$

where $<\mu_1> = J_1/\varphi_1$ is the mean cosine of the radiance at the boundary [11]. Inserting φ(0) from eqn (29) in eqns (33a,b) yields

$$A_2+B_2 = A_1(n_{21}^2+C_0(n_{21})D_1\lambda_1)+B_1(n_{21}^2-C_0(n_{21})D_1\lambda_1)$$
$$D_1\lambda_1[A_1-B_1] = D_2\lambda_2[A_2-B_2] \qquad (35a,b)$$

allowing $A_2$, $B_2$ to be expressed in terms of $A_1$, $B_1$ as follows

$$A_2 = \tfrac{1}{2}[n_{21}^2+C_0(n_{21})D_1\lambda_1+\frac{D_1\lambda_1}{D_2\lambda_2}]A_1$$
$$+\tfrac{1}{2}[n_{21}^2-C_0(n_{21})D_1\lambda_1-\frac{D_1\lambda_1}{D_2\lambda_2}]B_1 \qquad (36a)$$

$$B_2 = \tfrac{1}{2}[n_{21}^2+C_0(n_{21})D_1\lambda_1-\frac{D_1\lambda_1}{D_2\lambda_2}]A_1$$
$$+\tfrac{1}{2}[n_{21}^2-C_0(n_{21})D_1\lambda_1+\frac{D_1\lambda_1}{D_2\lambda_2}]B_1 \qquad (36b)$$

Interface boundary equations are required for multi-layer analysis, in addition to those for the external surfaces, yielding e.g. a 4x4 matrix for the intensity coefficients of a double layer, analogous to eqn (16) for asymptotic RT theory. The multi-layer diffusion equations can be solved sequentially as above, expressing the coefficients $A_{k+1}$, $B_{k+1}$ of the (k+1)th layer in terms of the coefficients $A_k$, $B_k$ of the preceding kth layer [see Appendix C]

**Results**

The albedo A* of a half-space illuminated by an obliquely incident uni-directional plane wave source is presented in Table I for scattering albedo $\varpi$=0.99, showing the dependence of A* on the angle of incidence. The errors in the diffusion and asymptotic results are comparable (~2%-3%); a boundary correction reduces the error in the asymptotic result by more than an order of magnitude (~0.1%). The error in the diffusion result increases rapidly for lower scattering albedoes, reaching 100% for $\varpi$=0.5, while the asymptotic error remains below 10% (Figs 3 a-f). Results of calculations of the dependence of half-space albedo A* on scattering albedo $\varpi$ are shown in Tables II (a, b) for plane-wave and isotropic sources, the diffusion values increasing in error as $\varpi$ diminishes, the asymptotic values remaining within a few percent of the RT data for the lowest scattering albedoes (<3% for $\varpi$=0.3). The asymptotic error is significantly less for isotropic illumination, remaining within ±2% for the full range of scattering albedo $\varpi \in |0, 1|$, whereas the error for plane-wave illumination has a maximum value of 8% when $\varpi$=0.8. The error is reduced by an order of magnitude with a suitable boundary correction (K-integral) (Fig 4 a,b). Similar results are found for a diffusing bounday (Fig 4 c,d). Results for the albedo A* and diffuse transmittance T* of a thick slab (d=10, n=1.5) with isotropic illumination [12] are presented in Table III. Note the precise agreement of the diffusion and asymptotic values for $\varpi \equiv 1$ (to 6 significant figures). For $\varpi$<1, the diffusion error increases as $\varpi$ diminishes, exceeding 100% for $\varpi$<0.5, the asymptotic error having a maximum value <3% for $\varpi$~0.8-0.9, decreasing thereafter (Fig 4e). Calculations with a boundary correction (K-integrals) achieve remarkable accuracy in this case, differing from the RT data by just 2 units in the 5[th] decimal [13]. The asymptotic flux density jump at an interface shows closer agreement with the K-integral calculations [5, 13] than diffusion (Table IV, Fig 2). Results for double-layers with differing refractive index are presented in Tables V, VI and Fig 4f for isotropic illumination, with similar results, the diffusion error exceeding 100% for $\varpi$<0.5, the asymptotic error remaining below a maximum ~2% ($\varpi$=0.8). This is reduced by ~2 orders of magnitude with a suitable boundary correction, as above. However, plots of the internal intensity profiles incident at a refracting boundary increasingly depart from the accurate RT profile, particularly for $\mu<\mu_c$ i.e. for total internal reflection (Supplement: Boundary flux). While this has little effect on the transmitted light (or the albedo), it does contribute to a significant error in the internal scalar flux (radiance) $\varphi$ at the boundary, even with the K-integral correction. Again intensity profiles for isotropic illumination are much closer to the RT profile, resulting in more accurate results, as above.

**Conclusions**

From the examples given comparing the performance of asymptotic radiative transfer and diffusion theory, it is clear that the results converge for particle scattering albedo $\varpi \Rightarrow 1$ as the angular intensities become increasingly isotropic, and that asymptotic RT is no more accurate than diffusion theory for scattering albedoes $\varpi \geq 0.9$, errors of $\approx 3\%-5\%$ occurring in calculations of half-space and slab albedo A* for either. However, diffusion theory diverges increasingly rapidly for lower scattering albedoes, the error exceeding 100% for $\varpi<0.5$, while the error in asymptotic RT remains below 10% for plane-wave illumination, passing through a broad maximum for $\varpi \approx 0.8$ and diminishing thereafter. Errors are lower for an isotropic source, the maximum albedo error $\Delta A^* \approx 3\%$ for $\varpi \approx 0.8$, falling below 1% for $\varpi<0.5$. Since both diffusion and asymptotic RT satisfy the same small number of boundary conditions viz. one for an external surface, two for an interface, four for a double layer, the latter is clearly advantageous for approximate radiative transfer calculations of albedo and diffuse transmittance for media with lower particle scattering albedoes ($\varpi \approx 0.8$ to $\varpi \Rightarrow 0$). More accurate results can be achieved by the addition of a suitable boundary correction to the asymptotic angular intensity profile via the method of K-integrals [5].

Asymptotic analysis employs only low-order matrices with elements determined by the scattering albedo and refractive index, which can be calculated to high accuracy by gaussian quadrature and conveniently tabulated (Appendix C)

**Appendix A**

The asymptotic angular intensity distribution g(μ,λ) is the solution of the characteristic equation of radiative transfer valid in the depth of the scattering medium [5]

$$(1-\lambda\mu)g(\mu,\lambda) = \varpi \int_{-1}^{1} d\mu' f(\mu' \to \mu) g(\mu',\lambda) \quad (A1)$$

where λ is the extinction coefficient (least eigenvalue of the characteristic equation), ϖ the particle scattering albedo and f(μ'→μ) the probability of scattering from μ' to μ, the direction cosines of the incident and scattered photon paths. For isotropic scattering, f(μ'→μ) ≡ ½ and

$$g(\mu,\lambda) = \frac{\varpi}{2(1-\lambda\mu)} \quad (A2)$$

and λ satisfies the normalisation [5]

$$1 = \frac{\varpi}{2\lambda} \log\left(\frac{1+\lambda}{1-\lambda}\right) \quad (A3)$$

For anisotropic scattering, g(μ,λ) is expanded as a series of Legendre polynomials $P_n(\mu)$ viz.

$$g(\mu,\lambda) = \sum_n (2n+1) b_n(\lambda) P_n(\mu) \quad (A4)$$

where the coefficients $b_n(\lambda)$ are found from the recursion relation [4]

$$\lambda(n+1)b_{n+1} - (2n+1)(1-\varpi g_n)b_n + \lambda b_{n-1} = 0 \quad (A5)$$

and the $g_n$ are the coefficients of the Legendre expansion of the phase function

$$p(\mu,\mu') = \sum_m (2m+1) g_m P_m(\mu) P_m(\mu') \quad (A6)$$

e.g. $g_n = g^n$ for Henyey-Greenstein scattering [14] and λ is the extinction coefficient (least eigenvalue of the characteristic equation (eqn (A1)). The eigenvalue λ may be calculated to any desired degree of accuracy from the recursion relation [14]

$$\Delta_n = \frac{(n\lambda)^2}{(2n+1)(1-\varpi g_n) - \Delta_{n+1}} \quad (A7)$$

with
$$1-\varpi = \Delta_1 \quad (A8)$$

Plots of the asymptotic intensity profiles for isotropic and Henyey-Greenstein scattering [15] are presented in Fig. A1.

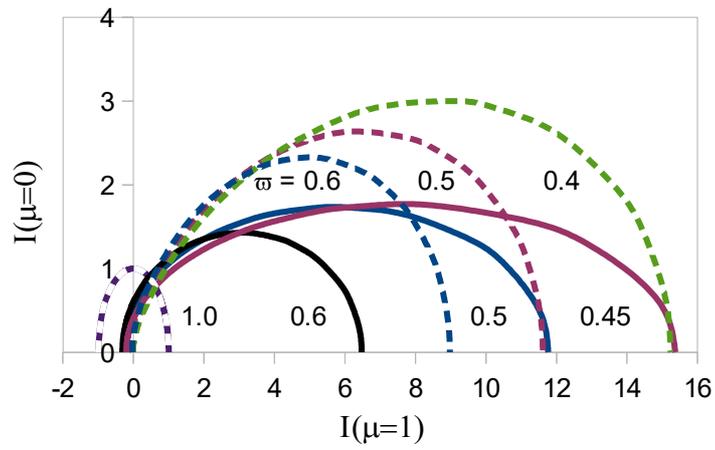

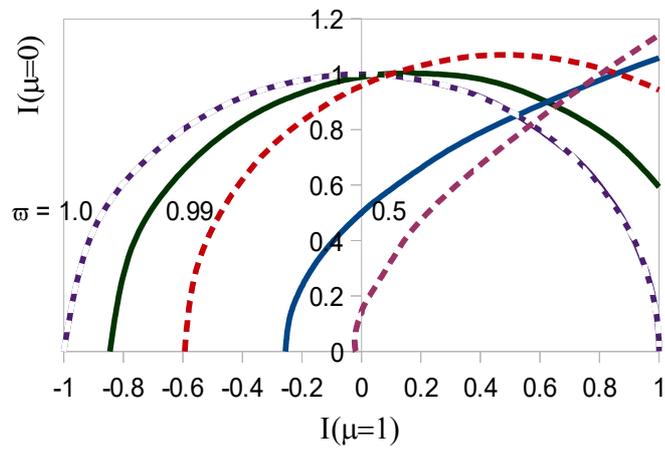

Fig A1

Polar plots of asymptotic radiance for isotropic (cont.) and Henyey-Greenstein (dashed) scattering (g=0.875) [15]. Curves labelled with value of scattering albedo $\varpi$

**Appendix B**

The asymptotic fluxes either side of the interface between two refractive half-spaces with constant uniform sources $Q_1$, $Q_2$ can be written [5]

$$I_1(x<0,\mu_1) = Q_1/2 - A_1 \exp(\lambda_1 x) g(\lambda_1, -\mu_1)$$
$$I_1(x>0,\mu_2) = Q_2/2 - A_2 \exp(-\lambda_2 x) g(\lambda_2, \mu_2) \quad \text{(B1a,b)}$$

and satisfy the boundary equations (cf eqns (15a,b))

$$\frac{Q_1}{2}[1 - R_{12}(\mu_1)] - A_1[g(\lambda_1, \mu_1) - R_{12}(\mu_1) g(\lambda_1, -\mu_1)]$$

$$= \frac{1}{n^2}[1 - R_{21}(\mu_2^-)]\left[\frac{Q_2}{2} - A_2 g(\lambda_2, -\mu_2^-)\right]$$

$$\frac{Q_2}{2}[1 - R_{21}(\mu_2)] - A_2[g(\lambda_2, \mu_2) - R_{21}(\mu_2) g(\lambda_2, -\mu_2)]$$

$$= n^2[1 - R_{12}(\hat{\mu}_1)]\left[\frac{Q_1}{2} - A_1 g(\lambda_1, -\hat{\mu}_1)\right] \quad \text{(B2a,b)}$$

the corresponding IBEs for $Q_1 = Q_2 = Q$ are

$$C(n_1) A_1 - \frac{F(n_2)}{n^2} A_2 = \frac{Q}{2}(1 - \frac{1}{n^2}) T_1$$

$$C(n_2) A_2 - n^2 F(n_1) A_1 = -\frac{Q}{2}(n^2 - 1) T_2 \quad \text{(B3a,b)}$$

with $C(n_i)$, $F(n_i)$ defined by eqns 12(a), 14(b) and $T_1$, $T_2$ are the transmission integrals

$$T_1 = \int_0^1 [1 - R_{12}(\mu_1)] \mu_1 d\mu_1$$

$$T_2 = \int_{\mu_c}^1 [1 - R_{21}(\mu_2)] \mu_2 d\mu_2 \quad \text{(B4a,b)}$$

The integrated boundary fluxes are

$$\Phi_1(0) = \frac{Q}{2} + A_1 \int_0^1 g(\lambda_1, -\mu_1) d\mu_1$$

$$\Phi_2(0) = \frac{Q}{2} + A_2 \int_0^1 g(\lambda_2, \mu_2) d\mu_2 \quad \text{(B5a,b)}$$

and the boundary flux discontinuity $\Delta\Phi(0) = \Phi_1(0) - \Phi_2(0)$ for $\lambda_1 = \lambda_2 = \lambda$

$$\Delta\Phi(0) = \frac{A_1 - A_2}{\lambda} \ln\left[\frac{1+\lambda}{1-\lambda}\right] = \frac{2(A_1 - A_2)}{\varpi} \quad \text{(B6)}$$

The coefficients $A_1$, $A_2$ are solutions of the 2x2 matrix equation (B3) with elements C1, C2, F1, F2, T1, T2 defined by eqns 12(a), 14(b) and eqn (B4) evaluated by gaussian quadrature and tabulated for n=1.333333 in Table B1. The coefficients $A_1$, $A_2$ and flux density difference $\Delta\Phi$ for a selection of scattering albedoes are presented in Table B2

**2x2 matrix elements**
**n=1.333333**

| ϖ | C1 | F2 | C2 | F1 | T2 | T1 |
|---|---|---|---|---|---|---|
| 1 | 0.466771 | 0.262559 | 0.262559 | 0.466771 | | |
| 0.9999 | 0.472848 | 0.258755 | 0.270304 | 0.461299 | 0.102106 | 0.102106 |
| 0.999 | 0.486511 | 0.250902 | 0.287470 | 0.449943 | | |
| 0.99 | 0.535004 | 0.229063 | 0.346169 | 0.417898 | | |
| 0.9 | 0.768772 | 0.181778 | 0.604712 | 0.345838 | | |
| 0.7 | 1.336585 | 0.154454 | 1.188857 | 0.302181 | | |
| 0.4 | 3.327631 | 0.143312 | 3.187069 | 0.283873 | | |

Table B1

Matrix elements for adjacent half-spaces with uniform internal sources calculated via 128-point gaussian quadrature (N1=N2=64)

| ϖ | | M | | | A | | A | T | | T | ΔΦ | asymp | K-int |
|---|---|---|---|---|---|---|---|---|---|---|---|---|---|
| 0.4 | **M** | 3.327631 | -0.143312 | **A** | **A1** | 0.029417 | **T** | **T1** | 0.102106 | ΔΦ | | | |
| | | -0.283873 | 3.187069 | | **A2** | -0.029417 | | **-T2** | -0.102106 | | 0.2942 | 0.3266 | |
| 0.7 | **M** | 1.336585 | -0.154454 | **A** | **A1** | 0.068480 | **T** | **T1** | 0.102106 | | | | |
| | | -0.302181 | 1.188857 | | **A2** | -0.068480 | | **-T2** | -0.102106 | | 0.3913 | 0.4058 | |
| 0.9 | **M** | 0.768772 | -0.181778 | **A** | **A1** | 0.107418 | **T** | **T1** | 0.102106 | | | | |
| | | -0.345838 | 0.604712 | | **A2** | -0.107418 | | **-T2** | -0.102106 | | 0.4774 | 0.4801 | |
| 0.99 | **M** | 0.535004 | -0.229063 | **A** | **A1** | 0.133635 | **T** | **T1** | 0.102106 | | | | |
| | | -0.417898 | 0.346169 | | **A2** | -0.133635 | | **-T2** | -0.102106 | | 0.5399 | 0.5386 | |
| 0.999 | **M** | 0.486511 | -0.250902 | **A** | **A1** | 0.138464 | **T** | **T1** | 0.102106 | | | | |
| | | -0.449943 | 0.287469 | | **A2** | -0.138468 | | **-T2** | -0.102106 | | 0.5544 | 0.5533 | |
| 0.9999 | **M** | 0.472848 | -0.258755 | **A** | **A1** | 0.139565 | **T** | **T1** | 0.102106 | | | | |
| | | -0.461299 | 0.270304 | | **A2** | -0.139565 | | **-T2** | -0.102106 | | 0.5583 | 0.5579 | |

Table B2

2x2 matrix solutions for adjacent half-spaces with uniform sources
ϖ – scattering albedo, **M** – matrix elements, **A** – intensity coefficients,
**T** – transmission integrals, ΔΦ – scalar flux discontinuity, comparing
asymptotic and K-integral results [5]

**Appendix C**

<u>Interface ($x_k = a_k = -a_{k+1}$)</u>
<u>Multilayer diffusion BEs (boundary equations)</u>

$$A_{k+1}\exp(\lambda_{k+1}a_{k+1}) + B_{k+1}\exp(-\lambda_{k+1}a_{k+1}) \qquad \varphi_{k+1} = n_{k+1}{}^2\varphi_k + C_0(n_{k+1})J_k$$
$$= A_k\exp(-\lambda_k a_k)(n_{k+1}{}^2 + C_0(n_{k+1})D_k\lambda_k) + B_k\exp(\lambda_k a_k)(n_{k+1}{}^2 - C_0(n_{k+1})D_k\lambda_k)$$
(C1a)

$$D_k\lambda_k[A_k\exp(-\lambda_k a_k) - B_k\exp(\lambda_k a_k)] \qquad J_k(a_k) = J_{k+1}(-a_{k+1})$$
$$= D_{k+1}\lambda_{k+1}[A_{k+1}\exp(\lambda_{k+1}a_{k+1}) - B_{k+1}\exp(-\lambda_{k+1}a_{k+1})] \quad J(a) = -D\nabla\varphi(a)$$
(C1b)

<u>Multilayer asymptotic IBEs (integrated boundary equations)</u>

$$A_{k+1}\exp(\lambda_{k+1}a_{k+1})C_{k+1}(n_{k+1},\lambda_{k+1}) + B_{k+1}\exp(-\lambda_{k+1}a_{k+1})D_{k+1}(n_{k+1},\lambda_{k+1})$$
$$= A_k\exp(-\lambda_k a_k)E_k(n_k,\lambda_k) + B_k\exp(\lambda_k a_k)F_k(n_k,\lambda_k)$$
(C2a)

$$A_k\exp(-\lambda_k a_k)D_k(n_k,\lambda_k) + B_k\exp(\lambda_k a_k)C_k(n_k,\lambda_k)$$
$$= A_{k+1}\exp(\lambda_{k+1}a_{k+1})F_{k+1}(n_{k+1},\lambda_{k+1}) + B_{k+1}\exp(-\lambda_{k+1}a_{k+1})E_{k+1}(n_{k+1},\lambda_{k+1})$$
(C2b)

<u>Multilayer diffusion solutions</u>

$$A_{k+1}" = \tfrac{1}{2}[n_{k+1}{}^2 + C_0(n_{k+1})D_k\lambda_k + \frac{D_k\lambda_k}{D_{k+1}\lambda_{k+1}}]A_k'$$
$$+ \tfrac{1}{2}[n_{k+1}{}^2 - C_0(n_{k+1})D_k\lambda_k - \frac{D_k\lambda_k}{D_{k+1}\lambda_{k+1}}]B_k'$$
(C3a)

$$B_{k+1}" = \tfrac{1}{2}[n_{k+1}{}^2 + C_0(n_{k+1})D_k\lambda_k - \frac{D_k\lambda_k}{D_{k+1}\lambda_{k+1}}]A_k'$$
$$+ \tfrac{1}{2}[n_{k+1}{}^2 - C_0(n_{k+1})D_k\lambda_k + \frac{D_k\lambda_k}{D_{k+1}\lambda_{k+1}}]B_k'$$
(C3b)

<u>Multilayer asymptotic solutions</u>

$$A_{k+1}" = \frac{(D_{k+1}D_k - E_{k+1}E_k)A_k' + (D_{k+1}C_k - E_{k+1}F_k)B_k'}{\det M_{k+1}}$$

$$B_{k+1}" = -\frac{(F_{k+1}E_k - C_{k+1}D_k)A_k' + (F_{k+1}F_k - C_{k+1}C_k)B_k'}{\det M_{k+1}}$$

$$A_{k+1}" = A_{k+1}\exp(\lambda_{k+1}a_{k+1}) \quad B_{k+1}" = B_{k+1}\exp(-\lambda_{k+1}a_{k+1})$$

$$A_k' = A_k\exp(-\lambda_k a_k) \quad B_k' = B_k\exp(\lambda_k a_k)$$

$$\det M_{k+1} = F_{k+1}D_{k+1} - E_{k+1}C_{k+1}, \ n = n_{k+1}/n_k$$

N.B. $\det M_{k+1} \equiv 0 \ (\varpi = 1)$

(C4)

## Interface asymptotic coefficients (isotropic scattering)

n=1.41(tissue)/1.3348(1% saline)=1.0563

| ϖ | C2 | E2 | D2 | F2 | C1 | E1 | D1 | F1 |
|---|---|---|---|---|---|---|---|---|
| 1 | 0.441182 | 0.441182 | 0.441182 | 0.441182 | 0.492292 | 0.492292 | 0.492292 | 0.492292 |
| 0.995 | 0.487710 | 0.484307 | 0.402083 | 0.405486 | 0.537366 | 0.536970 | 0.454746 | 0.455141 |
| 0.99 | 0.509615 | 0.504805 | 0.387699 | 0.392509 | 0.558698 | 0.558138 | 0.441033 | 0.441593 |
| 0.98 | 0.543705 | 0.536912 | 0.368932 | 0.375725 | 0.592012 | 0.591219 | 0.423238 | 0.424031 |
| 0.97 | 0.572680 | 0.564372 | 0.355669 | 0.363977 | 0.620415 | 0.619442 | 0.410739 | 0.411712 |
| 0.96 | 0.599270 | 0.589691 | 0.345189 | 0.354769 | 0.646542 | 0.645416 | 0.400915 | 0.402040 |
| 0.95 | 0.624532 | 0.613839 | 0.336455 | 0.347148 | 0.671407 | 0.670147 | 0.392763 | 0.394023 |
| 0.9 | 0.744145 | 0.729146 | 0.306211 | 0.321210 | 0.789588 | 0.787791 | 0.364856 | 0.366654 |
| 0.8 | 0.997406 | 0.976624 | 0.272807 | 0.293589 | 1.041168 | 1.038577 | 0.334760 | 0.337351 |
| 0.7 | 1.312770 | 1.287968 | 0.253564 | 0.278367 | 1.355527 | 1.352278 | 0.317874 | 0.321124 |
| 0.6 | 1.738614 | 1.710850 | 0.241340 | 0.269105 | 1.780729 | 1.776853 | 0.307344 | 0.311220 |
| 0.5 | 2.352288 | 2.322314 | 0.233550 | 0.263524 | 2.394004 | 2.389456 | 0.300692 | 0.305240 |
| 0.4 | 3.304258 | 3.272560 | 0.228801 | 0.260499 | 3.345755 | 3.340365 | 0.296605 | 0.301995 |
| 0.3 | 4.938022 | 4.904668 | 0.225897 | 0.259252 | 4.979428 | 4.972755 | 0.293984 | 0.300657 |

Table C1

n=1.333333

| ϖ | C2 | E2 | D2 | F2 | C1 | E1 | D1 | F1 |
|---|---|---|---|---|---|---|---|---|
| 1 | 0.262559 | 0.262559 | 0.262559 | 0.262559 | 0.466771 | 0.466771 | 0.466771 | 0.466771 |
| 1 | 0.270304 | 0.266478 | 0.254929 | 0.258755 | 0.472848 | 0.472388 | 0.460839 | 0.461299 |
| 0.999 | 0.28747 | 0.275367 | 0.238799 | 0.250902 | 0.486511 | 0.485054 | 0.448486 | 0.449943 |
| 0.995 | 0.320125 | 0.293008 | 0.210783 | 0.2379 | 0.513175 | 0.509907 | 0.427682 | 0.43095 |
| 0.99 | 0.346169 | 0.307722 | 0.190617 | 0.229063 | 0.535004 | 0.530364 | 0.413258 | 0.417898 |
| 0.95 | 0.476109 | 0.388421 | 0.111037 | 0.198725 | 0.649513 | 0.638791 | 0.361407 | 0.372128 |
| 0.9 | 0.604712 | 0.477751 | 0.054817 | 0.181778 | 0.768772 | 0.752941 | 0.330006 | 0.345838 |
| 0.8 | 0.867882 | 0.680633 | -0.023184 | 0.164065 | 1.021546 | 0.996921 | 0.293104 | 0.317729 |
| 0.7 | 1.188857 | 0.951665 | -0.082738 | 0.154454 | 1.336585 | 1.302906 | 0.268503 | 0.302181 |
| 0.6 | 1.618168 | 1.337555 | -0.131955 | 0.148659 | 1.762209 | 1.71785 | 0.24834 | 0.292699 |
| 0.5 | 2.233952 | 1.915179 | -0.173585 | 0.145187 | 2.375742 | 2.317413 | 0.228649 | 0.286978 |
| 0.4 | 3.187069 | 2.832895 | -0.210862 | 0.143312 | 3.327631 | 3.248963 | 0.205205 | 0.283873 |
| 0.3 | 4.821307 | 4.426654 | -0.252113 | 0.14254 | 4.96136 | 4.848931 | 0.170164 | 0.282593 |

Table C2

Tables of asymptotic coefficients for Fresnel interface calculated via 128-point gaussian quadrature (N1=N2=64)

**Half-space albedo vs incidence angle**
**Diffusion, asymptotic, K-integral, RT data**
**n=1.333 $\varpi$=0.99**

| $\theta$ deg | $\alpha_{diff}$ diff. | $\alpha_{asymp}$ asymp | $\alpha_{K2}$ K-int | $\alpha_T$ H-int | $\Delta\alpha/\alpha$ diff.% | $\Delta\alpha/\alpha$ asy% | $\Delta\alpha/\alpha$ K-int% |
|---|---|---|---|---|---|---|---|
| 0 | 0.6667 | 0.6686 | 0.6526 | 0.6519 | 2.28 | 2.56 | 0.11 |
| 5 | 0.6667 | 0.6686 | 0.6527 | 0.6521 | 2.25 | 2.53 | 0.09 |
| 10 | 0.6667 | 0.6686 | 0.6533 | 0.6527 | 2.15 | 2.44 | 0.09 |
| 15 | 0.6668 | 0.6686 | 0.6542 | 0.6536 | 2.01 | 2.29 | 0.09 |
| 20 | 0.6668 | 0.6687 | 0.6554 | 0.655 | 1.80 | 2.09 | 0.06 |
| 25 | 0.6669 | 0.6688 | 0.6570 | 0.6567 | 1.55 | 1.84 | 0.05 |
| 30 | 0.6671 | 0.669 | 0.6590 | 0.6588 | 1.26 | 1.55 | 0.03 |
| 35 | 0.6675 | 0.6693 | 0.6615 | 0.6613 | 0.93 | 1.21 | 0.03 |
| 40 | 0.6682 | 0.67 | 0.6644 | 0.6644 | 0.56 | 0.84 | 0.00 |
| 45 | 0.6693 | 0.6712 | 0.6680 | 0.6681 | 0.18 | 0.46 | -0.01 |
| 50 | 0.6713 | 0.6731 | 0.6725 | 0.6727 | -0.21 | 0.06 | -0.03 |
| 55 | 0.6746 | 0.6764 | 0.6783 | 0.6786 | -0.59 | -0.32 | -0.04 |
| 60 | 0.6801 | 0.6819 | 0.6863 | 0.6866 | -0.94 | -0.68 | -0.04 |
| 65 | 0.6895 | 0.6912 | 0.6977 | 0.6981 | -1.24 | -0.99 | -0.06 |
| 70 | 0.7052 | 0.7069 | 0.7150 | 0.7154 | -1.42 | -1.19 | -0.06 |
| 75 | 0.7321 | 0.7336 | 0.7424 | 0.7428 | -1.44 | -1.24 | -0.05 |
| 80 | 0.7782 | 0.7794 | 0.7877 | 0.788 | -1.24 | -1.09 | -0.04 |
| 85 | 0.8583 | 0.8591 | 0.8648 | 0.865 | -0.77 | -0.68 | -0.02 |
| 90 | 1.0000 | 1.0000 | 1.0000 | 1.0000 | 0.00 | 0.00 | 0.00 |

Table I

Half-space albedo $\alpha$ vs angle of incidence. Plane-wave surface source (n=1.333 $\varpi$=0.99). Comparison of diffusion, asymptotic and K-integral values vs radiative transfer (H-int) [8]

### planar

| ϖ | $\alpha_{diff}$ diffusion | $\alpha_s$ | $\alpha_{asy}$ asymp. | $\alpha_{K2}$ K-int. | $\alpha_T$ H-int. |
|---|---|---|---|---|---|
| 0.995 | 0.7476 | *0.8114* | 0.7484 | 0.7357 | 0.7352 |
| 0.99  | 0.6667 | 0.6657  | 0.6686 | 0.6526 | 0.6519 |
| 0.98  | 0.5691 | 0.5667  | 0.5734 | 0.5543 |        |
| 0.97  | 0.5049 | 0.5009  | 0.5116 | 0.4911 |        |
| 0.96  | 0.4566 | 0.4510  | 0.4657 | 0.4447 |        |
| 0.95  | 0.4178 | 0.4106  | 0.4293 | 0.4081 | 0.4070 |
| 0.9   | 0.2926 | 0.2773  | 0.3149 | 0.2952 | 0.2940 |
| 0.8   | 0.1656 | 0.1346  | 0.2056 | 0.1910 |        |
| 0.7   | 0.0936 | 0.0473  | 0.1477 | 0.1375 | 0.1364 |
| 0.6   | 0.0444 | -0.0168 | 0.1104 | 0.1036 |        |
| 0.5   | 0.0077 | -0.0682 | 0.0841 | 0.0796 | 0.0792 |
| 0.4   | -0.0213 | -0.1115 | 0.0645 | 0.0615 |       |
| 0.3   | -0.0450 | -0.1488 | 0.0494 | 0.0472 | 0.0480 |
| 0.2   | -0.0649 | -0.1806 | 0.0389 | 0.0351 |       |
| 0.1   | -0.0820 | -0.2071 |        |        |        |
| 0     | -0.0969 | -0.2292 | 0.0204 | 0.0204 | 0.0204 |

### isotropic

| ϖ | $\alpha_{diff}$ diffusion | $\alpha_s$ | $\alpha_{asy}$ asymp. | $\alpha_{K2}$ K-int. | $\alpha_T$ H-int. |
|---|---|---|---|---|---|
| 0.995 | 0.7304 | 0.7912 | 0.7602 | 0.7584 | 0.7584 |
| 0.99  | 0.6533 | 0.6523 | 0.6842 | 0.6812 | 0.6813 |
| 0.98  | 0.5603 | 0.5580 | 0.5935 | 0.5891 |        |
| 0.97  | 0.4991 | 0.4953 | 0.5346 | 0.5293 |        |
| 0.96  | 0.4530 | 0.4477 | 0.4908 | 0.4851 |        |
| 0.95  | 0.4161 | 0.4092 | 0.4561 | 0.4501 | 0.4501 |
| 0.9   | 0.2967 | 0.2822 | 0.3471 | 0.3408 | 0.3409 |
| 0.8   | 0.1758 | 0.1462 | 0.2430 | 0.2383 |        |
| 0.7   | 0.1072 | 0.0630 | 0.1876 | 0.1849 | 0.1850 |
| 0.6   | 0.0603 | 0.0019 | 0.1522 | 0.1509 |        |
| 0.5   | 0.0253 | -0.0471 | 0.1271 | 0.1270 | 0.1271 |
| 0.4   | -0.0023 | -0.0884 | 0.1084 | 0.1091 |       |
| 0.3   | -0.0249 | -0.1239 | 0.0942 | 0.0953 | 0.0950 |
| 0.2   | -0.0440 | -0.1542 | *0.0868* | *0.0879* |    |
| 0.1   | -0.0603 | -0.1795 |        |        |        |
| 0     | -0.0745 | -0.2005 | 0.0187 | 0.0187 | 0.0187 |

Tables II a, b

Half-space albedo α vs scattering albedo ϖ, normally incident plane-wave and isotropic sources (n=1.333) [8, 13]. The column headed $\alpha_s$ is calculated using accurate values of the diffusion coefficient D (eq (27))

| ϖ | A* | T* | A* | T* | A* | T* | A* | T* |
|---|---|---|---|---|---|---|---|---|
| | diffusion | | asymptotic | | K-integral | | RT | |
| 1 | 0.79781 | 0.20219 | 0.79781 | 0.20219 | ***0.79447*** | ***0.20553***  | ***0.79445*** | ***0.20555*** |
| 0.9999 | 0.79523 | 0.20030 | 0.79524 | 0.20029 | 0.79190 | 0.20363 | | |
| 0.9995 | 0.78518 | 0.19299 | | | 0.78188 | 0.19629 | | |
| 0.999 | 0.77317 | 0.18442 | 0.77333 | 0.18431 | 0.76994 | 0.18766 | | |
| 0.995 | 0.69497 | 0.13234 | | | 0.69244 | 0.13507 | | |
| 0.99 | 0.62603 | 0.09260 | 0.62924 | 0.09221 | 0.62461 | 0.09487 | | |
| 0.98 | 0.53507 | 0.05116 | 0.54221 | 0.05097 | 0.53641 | 0.05283 | | |
| 0.97 | 0.47452 | 0.03127 | 0.48555 | 0.03123 | 0.47902 | 0.03256 | | |
| 0.96 | 0.42948 | 0.02040 | 0.44426 | 0.02045 | 0.43730 | 0.02144 | | |
| 0.95 | 0.39375 | 0.01394 | 0.41212 | 0.01405 | 0.40493 | 0.01479 | | |
| 0.9 | 0.28059 | 0.00316 | 0.31507 | 0.00330 | 0.30803 | 0.00355 | | |
| 0.8 | 0.16728 | 0.00045 | 0.22780 | 0.00052 | 0.22268 | 0.00057 | | |
| 0.7 | 0.10162 | 0.00012 | 0.18378 | 0.00016 | ***0.18036*** | ***0.00018*** | ***0.18034*** | ***0.00018*** |
| 0.6 | 0.05504 | 0.00005 | 0.15616 | 0.00008 | 0.15413 | 0.00008 | | |
| 0.5 | 0.01863 | 0.00003 | 0.13699 | 0.00005 | 0.13603 | 0.00004 | | |
| 0.4 | -0.01140 | 0.00002 | 0.12289 | 0.00004 | 0.12271 | 0.00002 | | |
| 0.3 | -0.03678 | 0.00002 | 0.11215 | 0.00004 | ***0.11247*** | ***0.00001*** | ***0.11245*** | ***0.00002*** |

Table III

Albedo A* and diffuse transmittance T* of dielectric slabs for isotropic surface illumination, isotropic scattering (n=1.5, d=10). Note near 5-figure agreement of K-integral values with RT data (bold italic) [12, 13]

| ϖ | diffusion ΔΦ | asympt. ΔΦ | K-int* ΔΦ | diff. δ% | asy. δ% |
|---|---|---|---|---|---|
| 1.0000 | 0.5600 | 0.5600 | 0.5600 | 0.00 | 0.00 |
| 0.9999 | 0.5583 | 0.5583 | 0.5579 | -0.07 | -0.07 |
| 0.999 | 0.5548 | 0.5544 | 0.5533 | -0.27 | -0.20 |
| 0.99 | 0.5438 | 0.5399 | 0.5386 | -1.0 | -0.2 |
| 0.98 | 0.5373 | 0.5298 | 0.5285 | -1.7 | -0.2 |
| 0.97 | 0.5323 | 0.5213 | 0.5203 | -2.3 | -0.2 |
| 0.96 | 0.5281 | 0.5138 | 0.5131 | -2.9 | -0.1 |
| 0.95 | 0.5245 | 0.5068 | 0.5067 | -3.5 | 0.0 |
| 0.9 | 0.5101 | 0.4774 | 0.4801 | -6.2 | 0.6 |
| 0.85 | 0.4989 | 0.4527 | | | |
| 0.8 | 0.4892 | 0.4306 | 0.4394 | -11.3 | 2.0 |
| 0.75 | 0.4804 | 0.4103 | | | |
| 0.7 | 0.4721 | 0.3913 | 0.4058 | -16.3 | 3.6 |
| 0.65 | 0.4642 | 0.3734 | | | |
| 0.6 | 0.4565 | 0.3562 | 0.3758 | -21.5 | 5.2 |
| 0.55 | 0.4490 | 0.3398 | | | |
| 0.5 | 0.4415 | 0.3249 | 0.3490 | -26.5 | 6.9 |
| 0.45 | 0.4340 | 0.3088 | | | |
| 0.4 | 0.4265 | 0.2942 | 0.3266 | -30.6 | 9.9 |

*mmrw [5]

Table IV

Interface flux density discontinuity ΔΦ vs ϖ (uniform sources, $n_2/n_1=4/3$)

Comparison of diffusion, asymptotic, K-integral results and relative error δ%

| ϖ | A* | T* | A* | T* | A* | T* | A* | T* |
|---|---|---|---|---|---|---|---|---|
| | diffusion | | asymptotic | | K-integral | | RT | |
| 1 | 0.77688 | 0.17541 | 0.81247 | 0.18753 | 0.80927 | 0.19073 | 0.80926 | 0.19075 |
| 0.999 | 0.75630 | 0.16177 | 0.79252 | 0.17188 | | | | |
| 0.99 | 0.62549 | 0.08694 | 0.66844 | 0.0888 | 0.66457 | 0.09148 | | |
| 0.95 | 0.39048 | 0.01467 | 0.45473 | 0.01418 | | | | |
| 0.9 | 0.26348 | 0.00352 | 0.3468 | 0.00339 | 0.34061 | 0.00366 | | |
| 0.8 | 0.12751 | 0.00053 | 0.2429 | 0.00054 | | | | |
| 0.7 | 0.04434 | 0.00015 | 0.18773 | 0.00017 | 0.18489 | 0.00019 | 0.18499 | 0.00019 |
| 0.6 | -0.01675 | 0.00007 | 0.15222 | 0.00008 | | | | |
| 0.5 | -0.06054 | | 0.12713 | 0.00005 | 0.12695 | 0.00004 | | |
| 0.4 | -0.10706 | | 0.10844 | 0.00004 | | | | |
| 0.3 | | | 0.09402 | 0.00004 | 0.09515 | 0.00001 | 0.095 | 0.00002 |

Table V

Double-layer albedo A*, diffuse transmittance T* vs scattering albedo ϖ

($n_1 = 1.333$  $n_2 = 1.5$  $2a = 10$)  isotropic source, isotropic scattering [12]

$n_1=1.333\ n_2=1.5\ 2a=10$

| $\varpi_1$ | $\varpi_2$ | A* | T* | A* | T* | A* | T* |
|---|---|---|---|---|---|---|---|
| | | asymptotic | | K-integral | | RT | |
| 1 | 0.7 | 0.69223 | 0.00387 | 0.68974 | 0.00374 | 0.68976 | 0.00374 |
| 1 | 0.3 | 0.68196 | 0.00192 | 0.67968 | 0.00036 | 0.67967 | 0.00108 |
| 1 | 1 | 0.81251 | 0.18749 | 0.80927 | 0.19073 | 0.80926 | 0.19075 |
| 0.7 | 1 | 0.18789 | 0.00403 | 0.18501 | 0.00437 | 0.18511 | 0.00441 |
| 0.3 | 1 | 0.09405 | 0.00142 | 0.09515 | 0.00034 | 0.09501 | 0.00109 |

$n_1=1.5\ n_2=1.333\ 2a=10$

| $\varpi_1$ | $\varpi_2$ | A* | T* | A* | T* | A* | T* |
|---|---|---|---|---|---|---|---|
| 1 | 1 | 0.80984 | 0.19016 | | | 0.80926 | 0.19075 |
| 0.7 | 0.7 | 0.18379 | 0.00018 | | | 0.18035 | 0.00019 |
| 0.3 | 0.3 | 0.11215 | 0.00004 | | | 0.11245 | 0.00002 |
| 0.7 | 1 | 0.1839 | 0.00325 | 0.1805 | 0.00414 | 0.18046 | 0.00374 |
| 1 | 0.7 | 0.65529 | 0.00509 | 0.65655 | 0.00517 | 0.65897 | 0.00441 |
| 0.3 | 1 | 0.11218 | 0.00121 | | | 0.11246 | 0.00108 |
| 1 | 0.3 | 0.63808 | 0.00234 | | | 0.64144 | 0.00109 |

$n_1=1.5\ n_2=1.5\ 2a=10$

| $\varpi_1$ | $\varpi_2$ | A* | T* | A* | T* | A* | T* |
|---|---|---|---|---|---|---|---|
| 1 | 1 | 0.79781 | 0.20219 | 0.79447 | 0.20553 | 0.79445 | 0.20555 |

$n_1=1.5\ n_2=3\ 2a=10$

| $\varpi_1$ | $\varpi_2$ | A* | T* | A* | T* | A* | T* |
|---|---|---|---|---|---|---|---|
| 0.3 | 0.3 | 0.11216 | 0.11247 | 0.11245 | 0.00002 | 0.00001 | 0.00002 |
| 0.3 | 0.7 | 0.11216 | | 0.11245 | 0.00002 | | 0.00003 |
| 0.3 | 1 | 0.11218 | 0.11247 | 0.11246 | 0.00145 | 0.00038 | 0.0012 |
| 0.7 | 0.3 | 0.18377 | | 0.18033 | 0.00005 | | 0.00005 |
| 0.7 | 0.7 | 0.18377 | | 0.18034 | 0.00005 | | 0.00009 |
| 0.7 | 1 | 0.18389 | 0.18046 | 0.18045 | 0.00416 | 0.00468 | 0.0047 |
| 1 | 0.3 | 0.64019 | 0.63634 | 0.63632 | 0.00121 | 0.00054 | 0.00094 |
| 1 | 0.7 | 0.64329 | 0.6409 | 0.63924 | 0.00123 | 0.00164 | 0.00178 |
| 1 | 1 | 0.77335 | 0.77029 | 0.77008 | 0.22665 | 0.22971 | 0.22991 |

$n_1=1\ n_2=1\ 2a=10$

| $\varpi_1$ | $\varpi_2$ | A* | T* | A* | T* | A* | T* |
|---|---|---|---|---|---|---|---|
| 0.995 | 0.8 | 0.77666 | 0.00655 | 0.77938 | 0.00608 | 0.77939 | 0.00609 |

Table VI

Double-layer albedo A*, diffuse transmittance T* vs scattering albedo $\varpi$, isotropic source, isotropic scattering [12]

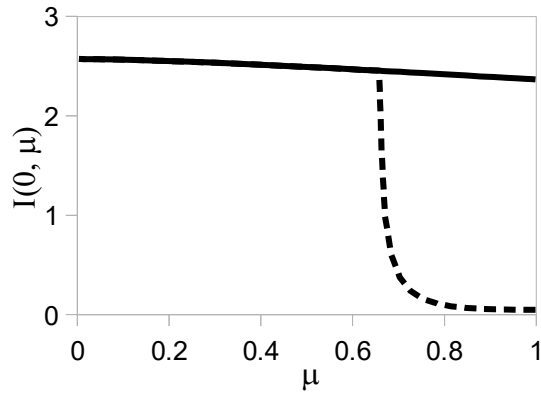

Fig. 1

Refractive 'hole' in reflected angular intensity at inner surface boundary (incident profile: continuous, reflected: dashed). Half-space (n = 1.333), plane-wave source, normal incidence, isotropic scattering, $\varpi=0.99$)

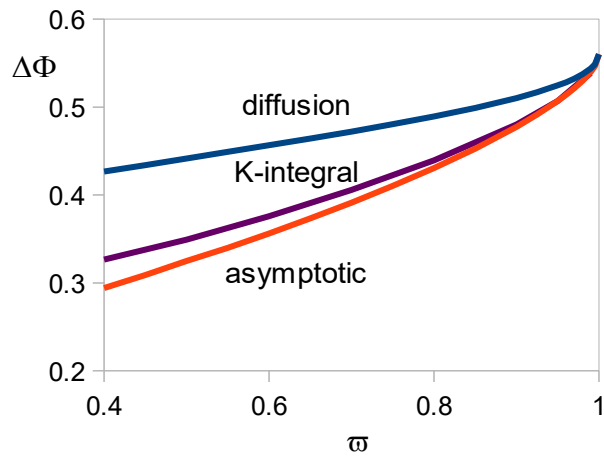

Fig 2

Comparison of diffusion, asymptotic transport and K-integral results for flux density discontuity at an interface between two scattering media of differing refractive index vs scattering albedo $\varpi$ (refractive index ratio n = 4/3, isotropic scattering) [5, 13]

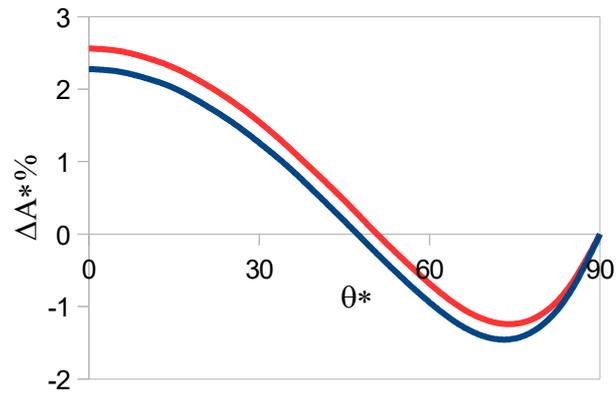

(a)   $\varpi=0.99$

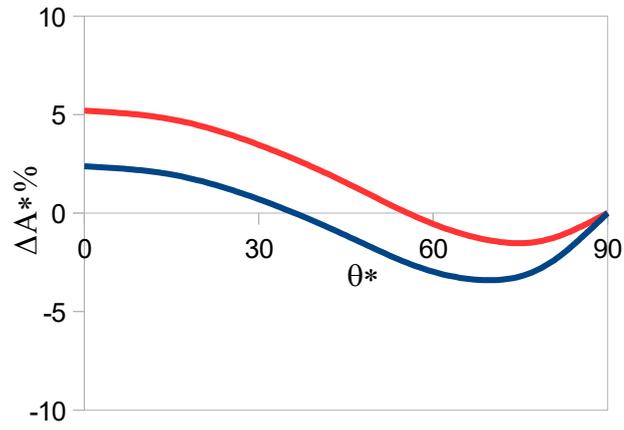

(b)   $\varpi=0.95$

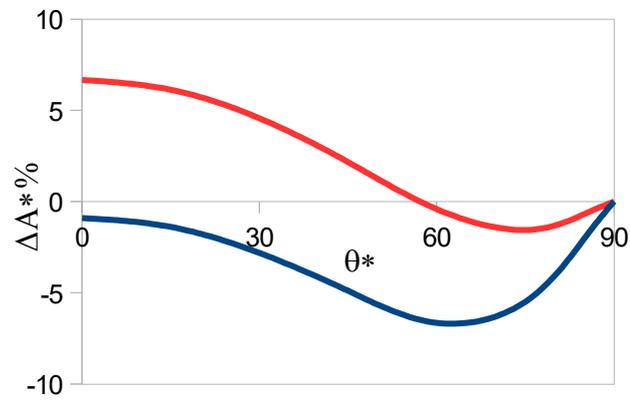

(c)   $\varpi=0.9$

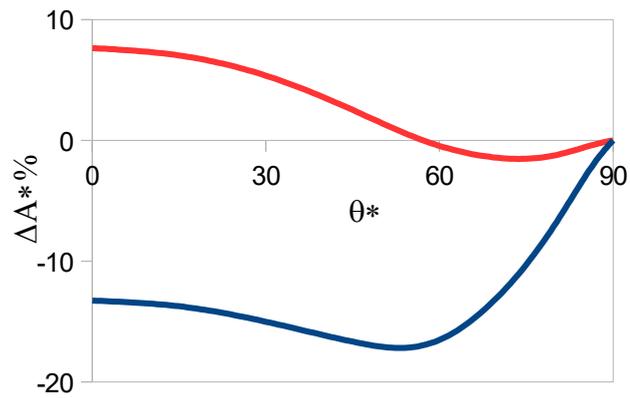

(d)   $\varpi=0.8$

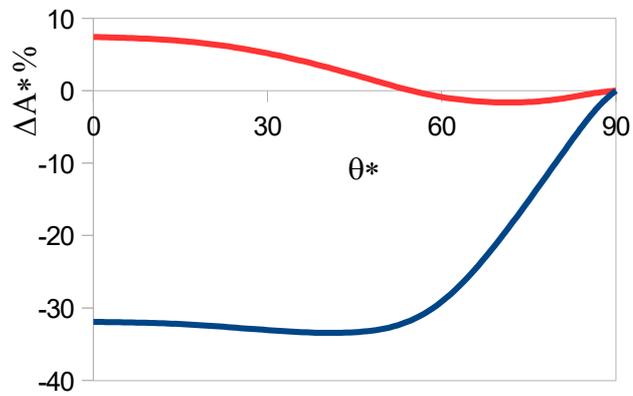

(e)   $\varpi=0.7$

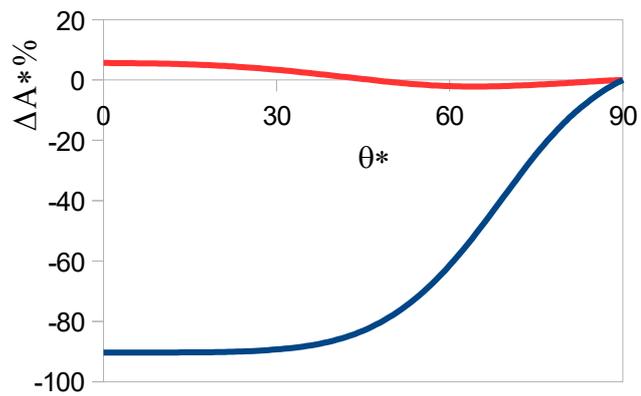

(f)   $\varpi=0.5$

Half-space albedo error $\Delta A^*\%$ vs plane wave angle of incidence $\theta*$ for a range of scattering albedo $0.99 \geq \varpi \geq 0.5$ (asymptotic - red, diffusion - blue, n=1.333)
Figs 3(a-f).

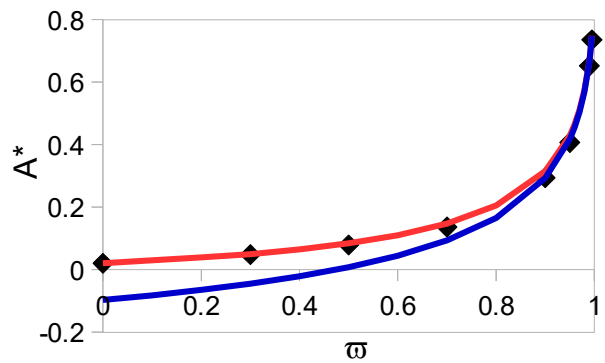

a) Half-space albedo A* vs scattering albedo $\varpi$. Plane-wave source, normal incidence, Fresnel boundary (n=1.333). Diffusion (blue), asymptotic (red), RT (♦)

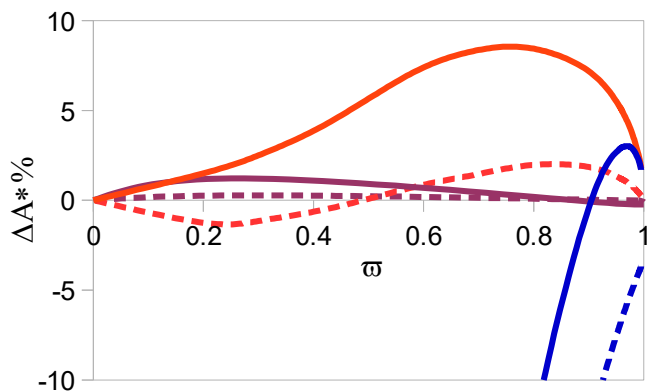

b) Half-space albedo error $\Delta A^*\%$ vs $\varpi$ : planar (cont.) and isotropic (dashed) sources. Diffusion (blue), asymptotic (red), K-integral (bordeaux) (n=1.333)

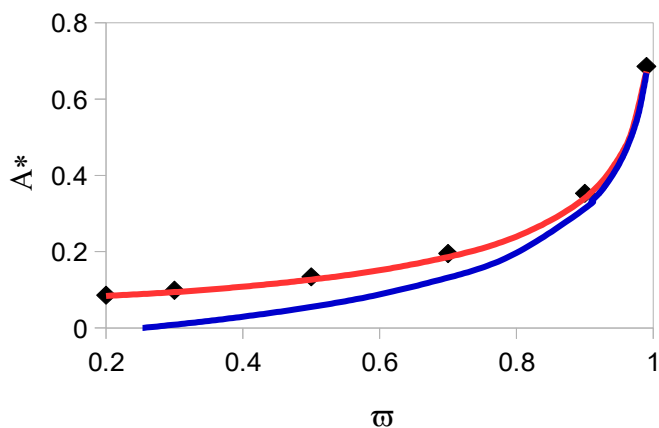

c) Half-space albedo A* vs scattering albedo $\varpi$. Plane-wave source (normal incidence), diffusing surface boundary (n=1.333). Asymptotic (red), diffusion (blue), RT (♦)

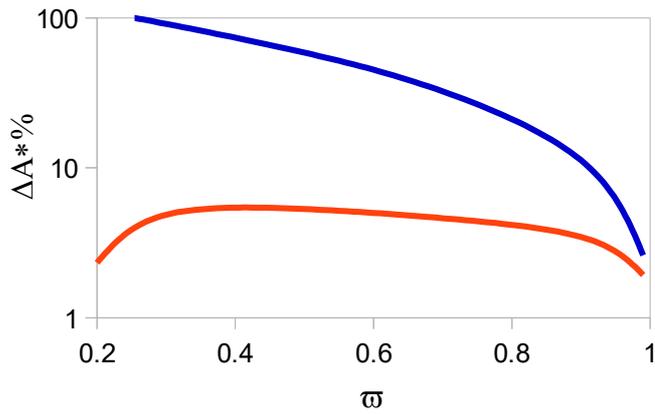

d) Half-space albedo error ΔA* % vs scattering albedo ϖ. Asymptotic (red), diffusion (blue), diffusing boundary. Parameters as above

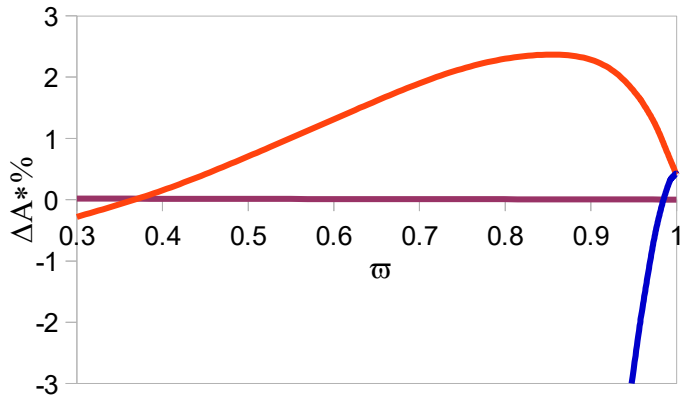

e) Slab albedo error ΔA*% vs scattering albedo ϖ (n=1.5 d=10)
Diffusion (blue), asymptotic (red), K-integral (bordeaux), isotropic source

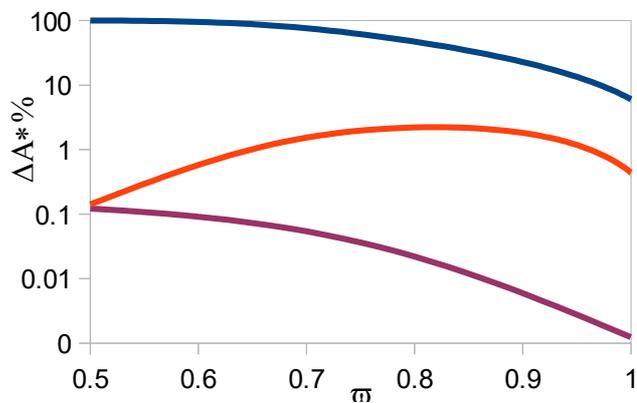

f) Double-layer albedo error |ΔA*|% vs ϖ: diffusion (blue), asymptotic (red), K-integral (bordeaux). Isotropic source ($n_1 = 1.333$, $n_2 = 1.5$, $2a = 10$)

Figs 4(a-f)

**Supplement: Boundary fluxes**

Further analysis of the angular intensity profiles corresponding to the diffusion, asymptotic and K-integral approximations reveals increasingly large errors in the internal intensity incident at the boundary $I(0, -\mu)$ for $\mu \leq \mu_c$ i.e. the region of total internal reflection, reaching a maximum error at $\mu=0$ (grazing incidence: $\theta=90$ deg). While having a small effect on external quantities, such as albedo $A^*$ and diffuse transmittance $T^*$ (comprising the transmitted light), this introduces significant errors in the internal scalar flux $\varphi$ (fluence) at the boundary, even for the K-integral analysis. The boundary error is significantly lower for an isotropic source compared with plane-wave illumination.

The boundary transient $\psi(0,\mu)$ used for the K-integral analysis takes the modified form

$$\psi(0,-\mu) \approx 1 - \beta\mu \ln\left(1 + \frac{1}{\mu}\right) \qquad (S1)$$

where $\beta \sim 1$ is a fitting parameter chosen to match the albedo $A^*$ given by RT theory [13]. However, the scalar flux error at the boundary, though reduced in this case, remains finite, requiring further development of the boundary transient to correct the error [5].

Results for half-space boundary fluxes are presented in the following for plane-wave and isotropic sources, comparing diffusion, asymptotic and K-integral predictions[‡] for $I(0,\mu)$, $J(0,\mu)$, $\varphi(0,\mu)$ and $\varphi(0)$ with accurate radiative transfer (RT) values [8]

---

[‡]The K-integrals are evaluated for $\psi(0,-\mu)$ constant (K1), log transient (as above) with $\beta=1$ (K2) and for $\beta=\beta'$ (K2') adjusted to fit the accurate RT albedo $A^*$. The integrals are calculated using gaussian quadrature, dividing the range of integration into two intervals either side of the cosine of the critical angle $\mu_c$ viz. $\mu \in |0, \mu_c|$ and $\mu \in |\mu_c, 1|$.

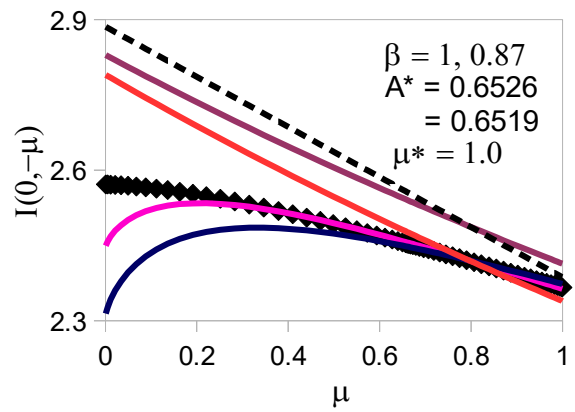

(a)

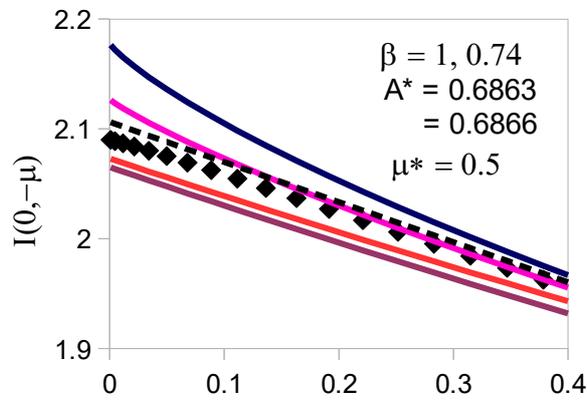

(b)

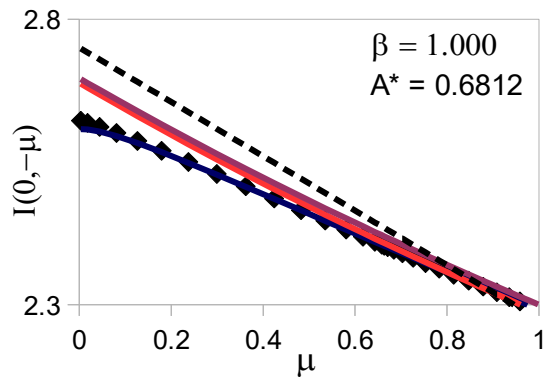

(c)

Internal intensity $I(0,-\mu)$ incident at the boundary of a half-space vs $\mu$ ($n = 1.333$, $\varpi = 0.99$). Diffusion (dashed), asymptotic (bordeaux), K-integral (K1: red, K2: blue, K2': pink), RT (♦). Plane-wave surface source incident at $\mu^* = 1$ (a), 0.5 (b); isotropic source (c)

Fig S1

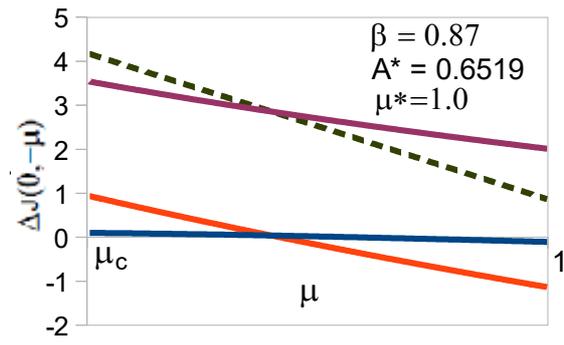

(a)

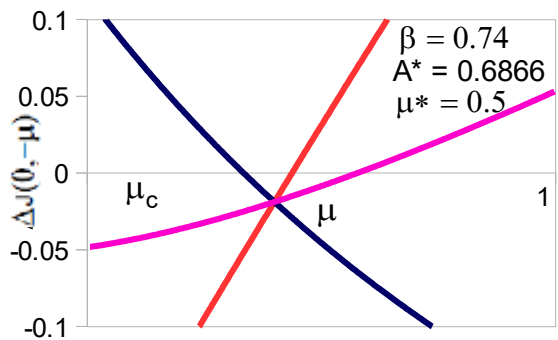

(b)

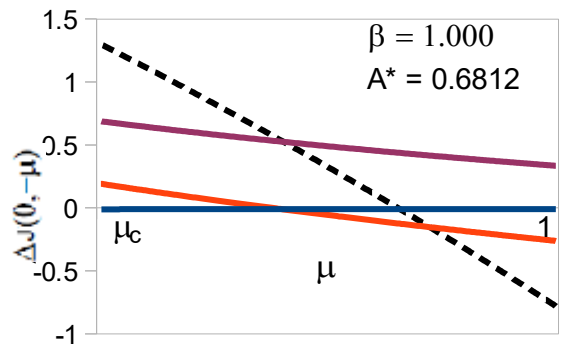

(c)

Percentage error $\Delta J(0,-\mu)$ at the boundary of a half-space vs $\mu$ ($n = 1.333$, $\varpi = 0.99$). Diffusion (dashed), asymptotic (bordeaux), K-integral (K1: red, K2: blue, K2': pink). Plane-wave surface source incident at $\mu^*=1$ (a), 0.5 (b); isotropic source (c)

Fig S2

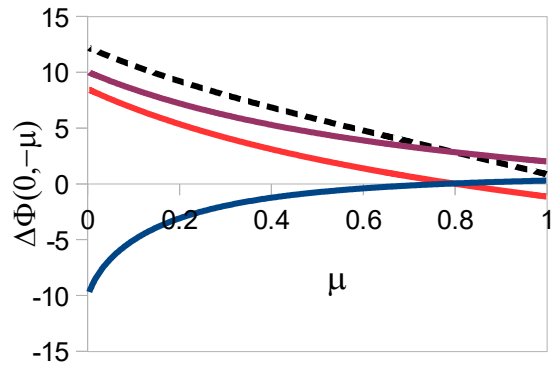

Percentage error ΔΦ(0,-μ) at the boundary of a half-space vs μ (n = 1.333, ϖ = 0.99).
Diffusion (dashed), asymptotic (bordeaux), K-integral (K1: red, K2: blue).
Plane-wave source at normal incidence

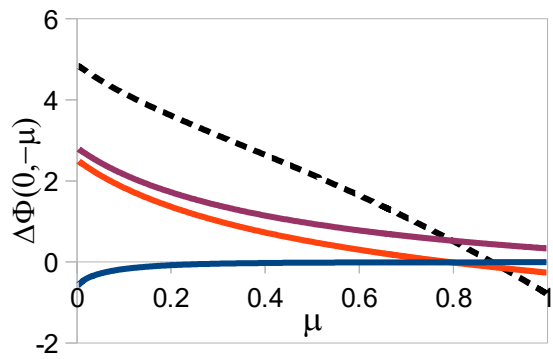

Percentage error ΔΦ(0,-μ) at the boundary of a half-space vs μ (n = 1.333, ϖ = 0.99).
Diffusion (dashed), asymptotic (bordeaux), K-integral (K1: red, K2: blue).
Isotropic surface source

Fig S3

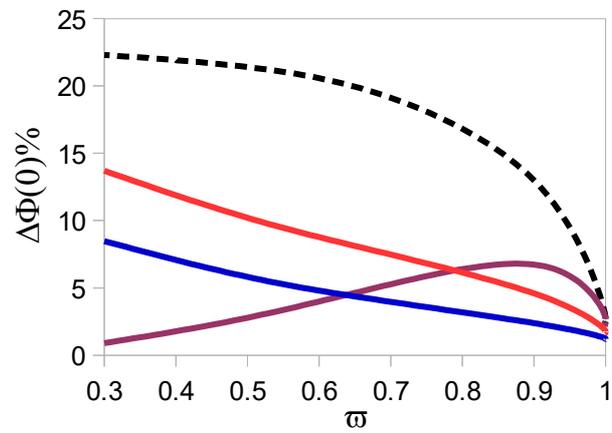

Half-space boundary flux error ΔΦ(0)% vs $\varpi$ (n=1.333333)
Planar source: diffusion (dashed), asymptotic (bordeaux),
K-integrals (K1: red, K2: blue)

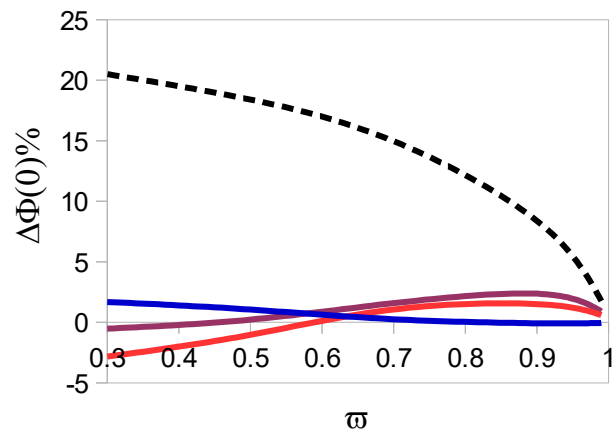

Half-space boundary flux error ΔΦ(0)% vs $\varpi$ (n=1.333333)
Lambertian source: diffusion (dashed), asymptotic (bordeaux),
K-integrals (K1: red, K2: blue)

Fig S4

## Half-space boundary fluxes: planar source, normal incidence, isotropic scattering n=4/3

| algorithm | $\Phi(0)$ | $\Delta\Phi(0)$ | J(0) | $\alpha(n,\nu)$ | $\beta'$, $\beta$ | |
|---|---|---|---|---|---|---|
| H-integral | 7.200788 | % | 0.84815 | 0.868558 | $\beta'$, $\beta$ | |
| K-integral2 | 7.178994 | -0.302654 | 0.84815 | 0.868558 | | 0.854 |
| K-integral2 | 7.116257 | -1.173914 | 0.848387 | 0.868795 | | 1 |
| K-integral1 | 7.316965 | 1.613395 | 0.84763 | 0.868038 | | |
| asymp | 7.356179 | 2.157974 | 0.854984 | 0.875392 | | |
| diffusion | 7.363664 | 2.261929 | 0.854911 | 0.875319 | | |

$\varpi=0.999$

| algorithm | $\Phi(0)$ | $\Delta\Phi(0)$ | J(0) | $\alpha(n,\nu)$ | $\beta'$, $\beta$ | |
|---|---|---|---|---|---|---|
| H-integral | 5.957014 | % | 0.631509 | 0.651917 | $\beta'$, $\beta$ | |
| K-integral2 | 5.935487 | -0.361373 | 0.63151 | 0.651918 | | 0.87 |
| K-integral2 | 5.872399 | -1.420434 | 0.632158 | 0.652566 | | 1 |
| K-integral1 | 6.094155 | 2.302171 | 0.62988 | 0.650288 | | |
| asymp | 6.188169 | 3.880381 | 0.648208 | 0.668616 | | |
| diffusion | 6.241073 | 4.768472 | 0.646345 | 0.666753 | | |

$\varpi=0.99$

| algorithm | $\Phi(0)$ | $\Delta\Phi(0)$ | J(0) | $\alpha(n,\nu)$ | $\beta'$, $\beta$ | |
|---|---|---|---|---|---|---|
| H-integral | 3.740997 | % | 0.273533 | 0.293942 | $\beta'$, $\beta$ | |
| K-integral2 | 3.716931 | -0.643321 | 0.273536 | 0.293944 | | 0.89 |
| K-integral2 | 3.651789 | -2.384616 | 0.274779 | 0.295187 | | 1.00 |
| K-integral1 | 3.912626 | 4.587766 | 0.2698 | 0.290208 | | |
| asymp | 3.99264 | 6.726612 | 0.29447 | 0.314879 | | |
| diffusion | 4.226983 | 12.99082 | 0.272147 | 0.292555 | | |

$\varpi=0.9$

| algorithm | $\Phi(0)$ | $\Delta\Phi(0)$ | J(0) | $\alpha(n,\nu)$ | $\beta'$, $\beta$ | |
|---|---|---|---|---|---|---|
| H-integral | 2.649565 | % | 0.115961 | 0.136369 | $\beta'$, $\beta$ | |
| K-integral2 | 2.628043 | -0.812276 | 0.115954 | 0.136362 | | 0.88 |
| K-integral2 | 2.544671 | -3.958924 | 0.117067 | 0.137475 | | 1.00 |
| K-integral1 | 2.847377 | 7.465806 | 0.113025 | 0.133433 | | |
| asymp | 2.789232 | 5.271308 | 0.127268 | 0.147676 | | |
| diffusion | 3.156256 | 19.12354 | 0.073216 | 0.093625 | | |

$\varpi=0.7$

| algorithm | $\Phi(0)$ | $\Delta\Phi(0)$ | J(0) | $\alpha(n,\nu)$ | $\beta'$, $\beta$ | |
|---|---|---|---|---|---|---|
| H-integral | 2.218628 | % | 0.058771 | 0.079179 | $\beta'$, $\beta$ | |
| K-integral2 | 2.314958 | 4.341849 | 0.05877 | 0.079179 | | 0.626 |
| K-integral2 | 2.089781 | -5.807536 | 0.059195 | 0.079604 | | 1 |
| K-integral1 | 2.444514 | 10.18133 | 0.058526 | 0.078934 | | |
| asymp | 2.280315 | 2.780388 | 0.063666 | 0.084074 | | |
| diffusion | 2.69383 | 21.41869 | -0.012698 | 0.00771 | | |

$\varpi=0.5$

# Half-space boundary fluxes: Lambertian source, isotropic scattering (n=4/3)

| algorithm | $\Phi(0)$ | $\Delta\Phi(0)$ | $J(0)$ | $\alpha(n,\nu)$ | $\beta', \beta$ |
|---|---|---|---|---|---|
| H-integral | 5.844054 | % | 0.614815 | 0.681274 | |
| <span style="color:red">K-integral2</span> | <span style="color:red">5.833747</span> | <span style="color:red">-0.176358</span> | <span style="color:red">0.614816</span> | <span style="color:red">0.681274</span> | <span style="color:red">1.067</span> |
| <span style="color:blue">K-integral2</span> | <span style="color:blue">5.841385</span> | <span style="color:blue">-0.045666</span> | <span style="color:blue">0.614737</span> | <span style="color:blue">0.681196</span> | <span style="color:blue">1</span> |
| K-integral1 | 5.879458 | 0.605817 | 0.614346 | 0.680805 | |
| asymp | 5.897265 | 0.910523 | 0.617736 | 0.684194 | |
| diffusion | 5.947689 | 1.773346 | 0.615961 | 0.68242 | |

$\varpi=0.99$

| algorithm | $\Phi(0)$ | $J(0)$ | $\alpha(n,\nu)$ | $\Delta\Phi(0)$ |
|---|---|---|---|---|
| H-integral | 4.43004 | 0.383638 | 0.450097 | % |
| K-integral2 | 4.426343 | 0.383609 | 0.450067 | -0.083446 |
| K-integral1 | 4.483636 | 0.382621 | 0.44908 | 1.209847 |
| asymp | 4.517684 | 0.38968 | 0.456139 | 1.978417 |
| diffusion | 4.670594 | 0.37874 | 0.445198 | 5.430068 |

$\varpi=0.95$

| algorithm | $\Phi(0)$ | $J(0)$ | $\alpha(n,\nu)$ | $\Delta\Phi(0)$ |
|---|---|---|---|---|
| H-integral | 3.716808 | 0.27442 | 0.340879 | % |
| K-integral2 | 3.714031 | 0.27439 | 0.340848 | -0.074708 |
| K-integral1 | 3.772936 | 0.273265 | 0.339724 | 1.510109 |
| asymp | 3.80484 | 0.280647 | 0.347106 | 2.368477 |
| diffusion | 4.028234 | 0.259383 | 0.325841 | 8.378861 |

$\varpi=0.9$

| algorithm | $\Phi(0)$ | $J(0)$ | $\alpha(n,\nu)$ | $\Delta\Phi(0)$ |
|---|---|---|---|---|
| H-integral | 2.616493 | 0.118538 | 0.184997 | % |
| K-integral2 | 2.623319 | 0.118434 | 0.184893 | 0.260871 |
| K-integral1 | 2.644444 | 0.118152 | 0.184611 | 1.068258 |
| asymp | 2.658079 | 0.121296 | 0.187754 | 1.589369 |
| diffusion | 3.007865 | 0.069787 | 0.136246 | 14.95787 |

$\varpi=0.7$

| algorithm | $\Phi(0)$ | $J(0)$ | $\alpha(n,\nu)$ | $\Delta\Phi(0)$ |
|---|---|---|---|---|
| H-integral | 2.168218 | 0.060591 | 0.12705 | % |
| K-integral2 | 2.191145 | 0.060507 | 0.126965 | 1.057417 |
| K-integral1 | 2.145857 | 0.060592 | 0.127051 | -1.031318 |
| asymp | 2.173105 | 0.060679 | 0.127137 | 0.225382 |
| diffusion | 2.567187 | -0.012095 | 0.054363 | 18.40076 |

$\varpi=0.5$

| algorithm | $\Phi(0)$ | $J(0)$ | $\alpha(n,\nu)$ | $\Delta\Phi(0)$ |
|---|---|---|---|---|
| H-integral | 1.905995 | 0.028582 | 0.095041 | % |
| K-integral2 | 1.938157 | 0.028823 | 0.095282 | 1.687414 |
| K-integral1 | 1.852 | 0.027911 | 0.094369 | -2.832862 |
| asymp | 1.896006 | 0.027706 | 0.094165 | -0.524044 |
| diffusion | 2.296827 | -0.062331 | 0.004127 | 20.50541 |

$\varpi=0.3$

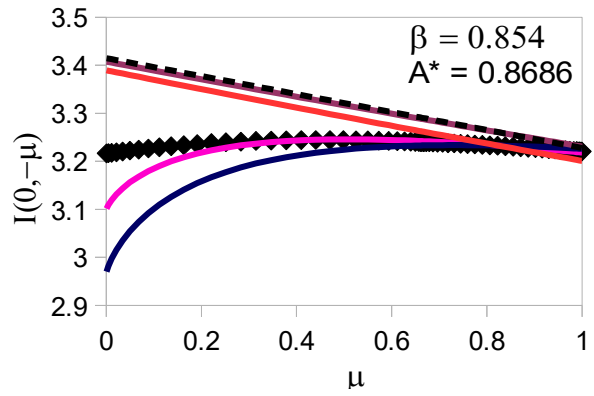

(a)

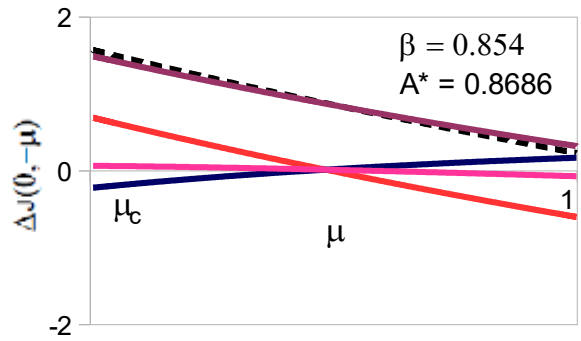

(b)

Fig S5

Half-space boundary values (plane-wave, normal incidence: n=1.333, $\varpi$=0.999, gaussian quadrature: N1=N2=32). Diffusion (dashed), asymptotic (bordeaux), K-integral (K1: red, K2: blue, K2': pink). (a) internal intensity $I(0, -\mu)$ vs $\mu$ (b) % error $\Delta J(0,-\mu)$ in boundary current $J(\mu)$ vs $\mu$ for $\mu \in |\mu_c, 1|$

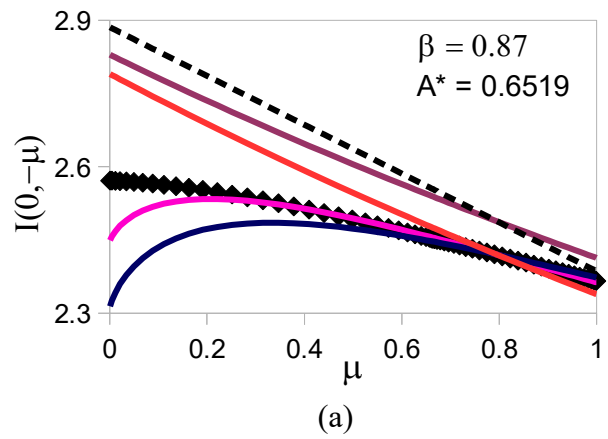

(a)

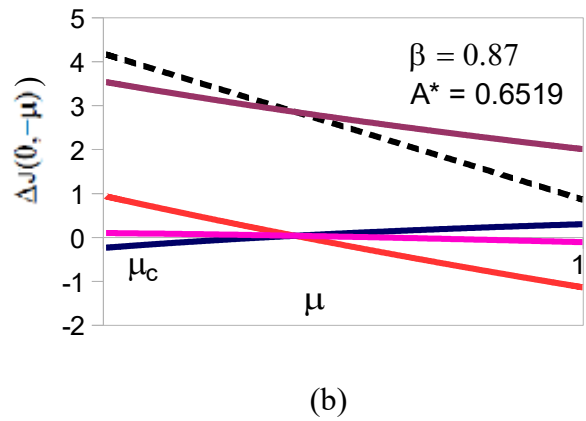

(b)

Fig S6

Half-space boundary values (plane-wave, normal incidence: n=1.333, ϖ=0.99, gaussian quadrature N1=N2=32). Diffusion (dashed), asymptotic (bordeaux), K-integral (K1: red, K2: blue, K2': pink). (a) internal intensity I(0, - μ) vs μ (b) % error ΔJ(0,-μ) in boundary current J(μ) vs μ for μ ∈ |μ$_c$,1|

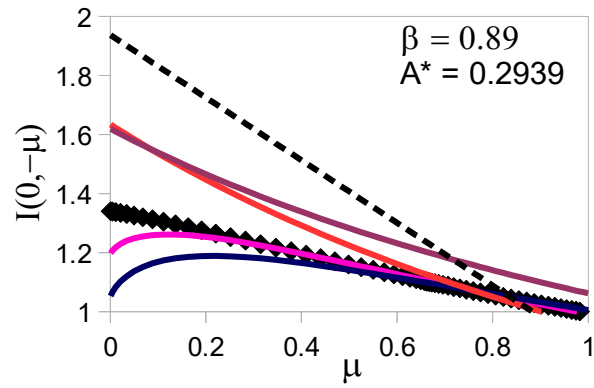

(a)

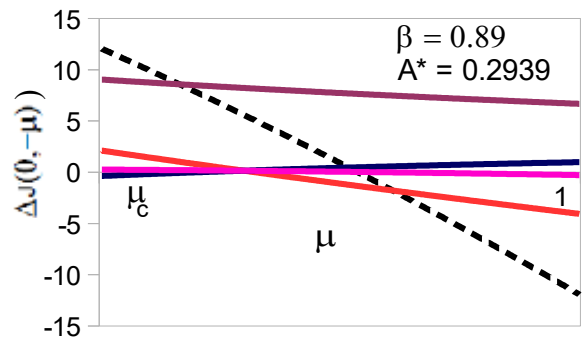

(b)

Fig S7

Half-space boundary values (plane-wave, normal incidence: n=1.333, $\varpi$=0.9, gaussian quadrature N1=N2=32). Diffusion (dashed), asymptotic (bordeaux), K-integral (K1: red, K2: blue, K2': pink). (a) internal intensity I(0, - μ) vs μ (b) % error ΔJ(0,-μ) in boundary current J(μ) vs μ for μ ∈ |μ_c,1|

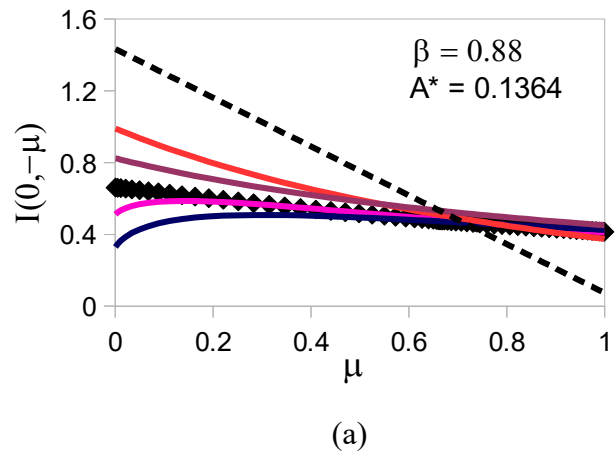

(a)

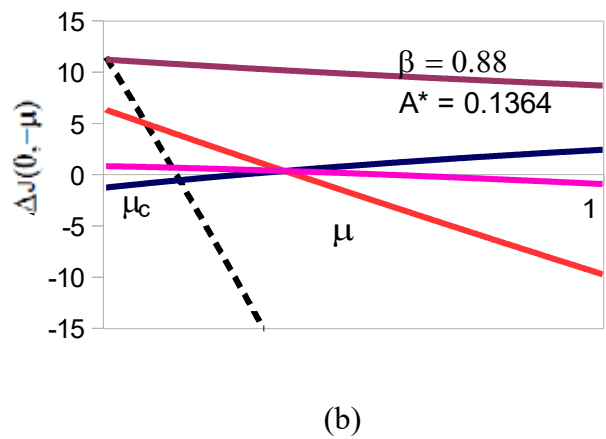

(b)

Fig S8

Half-space boundary values (plane-wave, normal incidence: n=1.333, $\varpi$=0.7, gaussian quadrature N1=N2=32). Diffusion (dashed), asymptotic (bordeaux), K-integral (K1: red, K2: blue, K2': pink). (a) internal intensity $I(0, -\mu)$ vs $\mu$ (b) % error $\Delta J(0,-\mu)$ in boundary current $J(\mu)$ vs $\mu$ for $\mu \in |\mu_c, 1|$

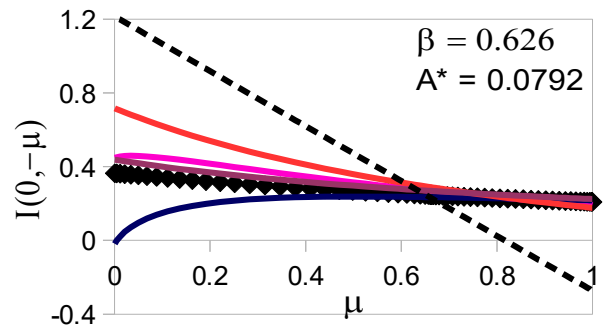

(a)

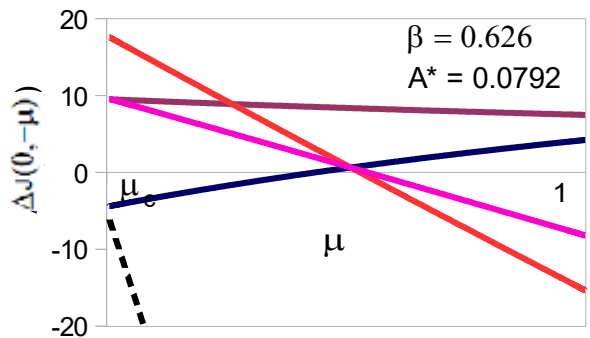

(b)

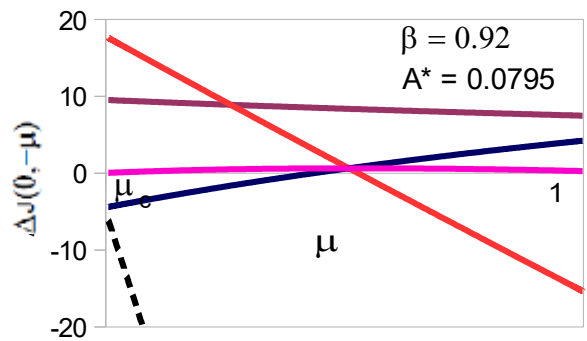

(c)

Fig S9

Half-space boundary values (plane-wave, normal incidence: n=1.333, $\varpi$=0.5, gaussian quadrature N1=N2=32). Diffusion (dashed), asymptotic (bordeaux), K-integral (K1: red, K2: blue, K2': pink). (a) internal intensity $I(0, -\mu)$ vs $\mu$ (b, c) % error $\Delta J(0,-\mu)$ in boundary current $J(\mu)$ vs $\mu$ for $\mu \in |\mu_c, 1|$